\documentclass[cernpreprint,texlive=2016,txfonts,texmf,USenglish]{atlasdoc}

\makeatletter
\DeclareOldFontCommand{\rm}{\normalfont\rmfamily}{\mathrm}
\DeclareOldFontCommand{\sf}{\normalfont\sffamily}{\mathsf}
\DeclareOldFontCommand{\tt}{\normalfont\ttfamily}{\mathtt}
\DeclareOldFontCommand{\bf}{\normalfont\bfseries}{\mathbf}
\DeclareOldFontCommand{\it}{\normalfont\itshape}{\mathit}
\DeclareOldFontCommand{\sl}{\normalfont\slshape}{\@nomath\sl}
\DeclareOldFontCommand{\sc}{\normalfont\scshape}{\@nomath\sc}
\makeatother
\usepackage[biblatex=false]{atlaspackage}
\usepackage{atlasbiblatex}
\usepackage{atlascontribute}
\usepackage{atlasphysics}

\usepackage{float} 
\usepackage{amssymb}
\usepackage{url}
\usepackage{harpoon}
\usepackage{amsmath}
\usepackage{MnSymbol}
\usepackage{multirow}
\usepackage{bm}
\usepackage{cite}

\newcommand{\lr}[1]{\left\langle #1\right\rangle}
\newcommand{\llrr}[1]{\left\llangle #1\right\rrangle}
\newcommand{\pPb}{\mbox{$p$+Pb}}

\newcommand{\nch}{\left\langle N_{\mathrm{ch}}\right\rangle}
\newcommand{\nchb}{\mbox{$N_{\mathrm{ch}}^{\mathrm{sel}}$}}
\newcommand{\nchc}{\mbox{$N_{\mathrm{ch}}^{\mathrm{rec}}$}}
\newcommand{\nchrec}{\mbox{$N_{\mathrm{ch}}^{\mathrm{rec}}$}}

\newcommand{\sqrtsnn}{\mbox{$\sqrt{s_{\mathrm{NN}}}$}}
\newcommand{\Dphi}{\mbox{$\Delta \phi$}}
\newcommand{\Deta}{\mbox{$\Delta \eta$}}

\AtlasTitle{Measurement of long-range multi-particle azimuthal correlations with the subevent cumulant method in $pp$ and $p$+Pb collisions with the ATLAS detector at the LHC}

\author{The ATLAS Collaboration}

\PreprintIdNumber{CERN-PH-2017-160}

\AtlasDate{\today}

\AtlasJournal{Phys.\ Rev. \ C.}

\AtlasAbstract{
A detailed study of multi-particle azimuthal correlations is presented using $pp$ data at $\sqrt{s}=5.02$ and 13 TeV, and $p$+Pb data at $\sqrt{s_{\rm{NN}}}=5.02$ TeV, recorded with the ATLAS detector at the LHC. The azimuthal correlations are probed using four-particle cumulants $c_{n}\{4\}$ and flow coefficients $v_n\{4\}=(-c_{n}\{4\})^{1/4}$ for $n=2$ and 3, with the goal of extracting long-range multi-particle azimuthal correlation signals and suppressing the short-range correlations. The values of $c_{n}\{4\}$ are obtained as a function of the average number of charged particles per event, $\left\langle N_{\rm{ch}} \right\rangle$, using the recently proposed two-subevent and three-subevent cumulant methods, and compared with results obtained with the standard cumulant method. The standard method is found to be strongly biased by short-range correlations, which originate mostly from jets with a positive contribution to $c_{n}\{4\}$. The three-subevent method, on the other hand, is found to be least sensitive to short-range correlations. The three-subevent method gives a negative $c_{2}\{4\}$, and therefore a well-defined $v_2\{4\}$, nearly independent of $\left\langle N_{\rm{ch}} \right\rangle$, which implies that the long-range multi-particle azimuthal correlations persist to events with low multiplicity. Furthermore, $v_2\{4\}$ is found to be smaller than the $v_2\{2\}$ measured using the two-particle correlation method, as expected for long-range collective behavior. Finally, the measured values of $v_2\{4\}$ and $v_2\{2\}$ are used to estimate the number of sources relevant for the initial eccentricity in the collision geometry. The results based on the subevent cumulant technique provide direct evidence, in small collision systems, for a long-range collectivity involving many particles distributed across a broad rapidity interval.}
\hypersetup{pdftitle={ATLAS document},pdfauthor={The ATLAS Collaboration}}

\begin{document}

\maketitle

\section{Introduction}
\label{intro}

The study of azimuthal correlations in high-energy nuclear collisions at the Relativistic Heavy Ion Collider (RHIC) and the Large Hadron Collider (LHC) has been important for understanding the multi-parton dynamics of QCD in the strongly coupled non-perturbative regime. One striking observation is the long-range ``ridge''~\cite{Abelev:2009af,Alver:2009id,ALICE:2011ab,Aad:2012bu,Chatrchyan:2013kba} in two-particle angular correlations (2PC): an apparent collimated emission of particle pairs with small relative azimuthal angle ($\Dphi$) and large separation in pseudorapidity ($\Deta$). The ridge signature from 2PC is characterized by a Fourier decomposition of the correlation function $C(\Dphi)\sim 1+2\sum_n v_n^2\cos (n \Dphi)$, where $v_n$ denotes the single-particle anisotropy harmonic coefficients. The second-order coefficient $v_2$ is observed to be the largest, followed by $v_3$~\cite{ALICE:2011ab,Aad:2012bu}. These coefficients carry information about the collective behavior of the produced system. The ridge was first discovered in nucleus--nucleus (A+A) collisions~\cite{Adare:2008ae,Abelev:2009af,Alver:2009id,ALICE:2011ab,Aad:2012bu,Chatrchyan:2013kba}, but was later observed in small systems such as proton--nucleus ($p$+A) collisions~\cite{CMS:2012qk,Abelev:2012ola,Aad:2012gla,Aad:2014lta,Khachatryan:2015waa}, light-ion--nucleus collisions~\cite{Adare:2013piz}, and more recently in proton--proton ($\pp$) collisions~\cite{Khachatryan:2010gv,Aad:2015gqa,Aaboud:2016yar,atlas:4}. The ridge in large systems, such as central or mid-central A+A collisions, is commonly interpreted as the result of collective hydrodynamic expansion of hot and dense nuclear matter created in the overlap region of the colliding nuclei. Since the formation of an extended region of nuclear matter is not expected in small collision systems such as $p$+A and $\pp$, the origin of the ridge there could be different from that formed in large collision systems. There remains considerable debate in the theoretical community as to whether the ridge in small systems is of hydrodynamic origin, like it is in A+A collisions~\cite{Bozek:2013uha}, or stems from other effects such as initial-state gluon saturation~\cite{Dusling:2013qoz}.

An important question about the ridge is whether it involves all particles in the event (collective flow) or if it arises merely from correlations among a few particles, due to resonance decays, jets, or multi-jet production (non-flow). In small systems the contributions from non-flow sources, in particular from jets and dijets, are large. The extraction of a ridge signal using the 2PC method requires a large $\Delta\eta$ gap and careful removal of the significant contribution from dijet production~\cite{Abelev:2012ola,Aad:2012gla,Aad:2014lta,Aad:2015gqa,Aaboud:2016yar,Khachatryan:2016txc}. Since collective flow is intrinsically a multi-particle phenomenon, it can be probed more directly using cumulants based on multi-particle correlation techniques~\cite{Borghini:2000sa}. Azimuthal correlations involving four, six, and eight particles have been measured in $p$+Pb, d+Au and $\pp$ collisions, and a significant $v_2$ signal has been obtained~\cite{Aad:2013fja,Khachatryan:2015waa,Khachatryan:2016txc,Aidala:2017ajz}. One weakness of the standard multi-particle cumulant method is that it does not suppress adequately the non-flow correlations in small systems, which lead to a sign-change of $c_2\{4\}$ at smaller values of the charged particle multiplicity, $N_{\rm{ch}}$~\cite{Aad:2013fja,Khachatryan:2015waa,Khachatryan:2016txc,atlas:4}. Furthermore, the magnitude of $c_2\{4\}$ and the $N_{\rm{ch}}$ value at which the sign change occurs are found to depend sensitively on the exact definition of $N_{\rm{ch}}$ used to categorize the events. These observations suggest that the standard cumulant method, on which several previous measurements in small systems are based, is strongly contaminated by non-flow correlations~\cite{Aad:2013fja,Khachatryan:2015waa,Khachatryan:2016txc,Aidala:2017ajz}, especially in $\pp$ collisions and low $N_{\rm{ch}}$ region.

Recently an improved cumulant method based on the correlation between particles from different subevents separated in $\eta$ has been proposed to further reduce the non-flow correlations~\cite{jjia}. The effectiveness of this method for suppressing non-flow correlations has been validated using the $\PYTHIA{8}$ event generator~\cite{Sjostrand:2007gs}, which contains only non-flow correlations.

This paper presents measurements of $c_2\{4\}$ and $c_3\{4\}$ in $\pp$ collisions at $\sqrt{s}=$ 5.02 and 13~TeV, as well as $p$+Pb collisions at $\sqrtsnn=$ 5.02~TeV. They are obtained using two- and three-subevent cumulant methods and are compared with the standard cumulant method. The $c_2\{4\}$ cumulant is converted to the corresponding $v_2$ coefficient and compared with the results obtained using the two-particle correlation method in Refs.~\cite{Aaboud:2016yar,Aad:2014lta} to assess the nature of the event-by-event fluctuation of the collective flow in these collisions.

The paper is organized as follows. Section~\ref{sec:method} describes the framework for the standard, two-subevent and three-subevent four-particle cumulant methods used in this analysis. Details of the detector, trigger, datasets, as well as event and track selections are provided in Sections~\ref{sec:data} to~\ref{sec:sel}. The correlation analysis and systematic uncertainties are described in Sections~\ref{sec:corr} and~\ref{sec:syscheck}, respectively. The measured cumulants from the three datasets are provided in Section~\ref{sec:result}. A summary is given in Section~\ref{sec:sum}.

\section{Four-particle cumulants}
\label{sec:method}
The multi-particle cumulant method~\cite{Borghini:2000sa} is used to extract the amplitude of long-range azimuthal correlations of particles produced in high-energy collisions. This method has the advantage of suppressing correlations from jets and dijets, instead of relying on an explicit procedure to correct $v_n$ harmonics for dijet contributions in the 2PC approach, as done in Refs.~\cite{Aad:2014lta,Aad:2015gqa}. The framework for the standard cumulant is described in Refs.~\cite{Bilandzic:2010jr,Bilandzic:2013kga}, which was recently extended to the case of subevent cumulants in Ref.~\cite{jjia}.  This paper presents measurements of four-particle cumulants obtained with the standard, two-subevent and three-subevent methods. The following discussion first describes the standard cumulant method, then describes the two- and three-subevent methods focusing on the differences from the standard method.

The cumulant methods involve the calculation of $2k$-particle azimuthal correlations $\lr{\{2k\}_n}$, and $2k$-particle cumulants, $c_n\{2k\}$, for the $n^{\rm{th}}$-order flow harmonics. The two- or four-particle azimuthal correlations in one event are evaluated as~\cite{Bilandzic:2010jr,Bilandzic:2013kga,jjia}:
\begin{eqnarray}\label{eq:1}
\hspace*{-0.7cm}&&\lr{\{2\}_n} = \lr{{\mathrm{e}}^{{\rm i}n(\phi_1-\phi_2)}} = \frac{q_n^2-\tau_1}{1-\tau_1}\;,\\\label{eq:2}
\hspace*{-0.7cm}&&\lr{\{4\}_n} = \lr{{\mathrm{e}}^{{\rm i}n(\phi_1+\phi_2-\phi_3-\phi_4)}} = \frac{q_n^4-2\tau_1({\mathrm{Re}}[{\bm q}_{2n;2}{\bm q}_n^{*2}] +2q_{n}^2)+8\tau_2{{\mathrm{Re}}[{\bm q}_{n;3}}{\bm q}_n^*]+\tau_1^2(2+q_{2n;2}^2)-6\tau_3 }{1-6\tau_1+8\tau_2+3\tau_1^2-6\tau_3},\;\;\;\;\;
\end{eqnarray}
where ``$\lr{}$'' denotes a single-event average over all pairs or quadruplets, respectively. The averages from Eqs.~\eqref{eq:1} and \eqref{eq:2} are expanded into per-particle normalized flow vectors ${\bm q}_{n;l}$ and factors $\tau_l$ with $l=1,2...$~:
\begin{eqnarray}\label{eq:3}
{\bm q}_{n;l} \equiv \frac{\sum\limits_j  w_j^l{\mathrm{e}}^{{\rm i}n\phi_j}}{\sum\limits_jw_j^l}\;, q_{n;l}\equiv |{\bm q}_{n;l}|\;,\;q_{n}\equiv q_{n;1}\;,\;\tau_l \equiv \frac{\sum\limits_jw_j^{l+1}}{(\sum\limits_jw_j)^{l+1}}\;,
\end{eqnarray}
where the sum runs over all $M$ particles in the event and $w_j$ is a weight assigned to the $j^{\rm{th}}$ particle. This weight is constructed to correct for both detector non-uniformity and tracking inefficiency as explained in Section~\ref{sec:corr}. For unit weight $w_j=1$, then ${\bm q}_{mn;m}={\bm q}_{mn}$, and $\tau_l=1/M^l$. 

The two- and four-particle cumulants are obtained from the azimuthal correlations as:
\begin{eqnarray}\label{eq:4}
&&c_n\{2\}=\llrr{\{2\}_n}\;,\\\label{eq:4b}
&&c_n\{4\}=\llrr{\{4\}_n}-2\llrr{\{2\}_n}^2\;,
\end{eqnarray}
where ``$\llrr{}$'' represents a weighted average of $\lr{\{2k\}_n}$ over an event ensemble. In the absence of non-flow correlations, $c_n\{2k\}$ reflects the moments of the distribution of the flow coefficient $v_n$:
\begin{eqnarray}\label{eq:5}
c_n\{2\}_{\mathrm{flow}}=\lr{v_n^2}\;,\; c_n\{4\}_{\mathrm{flow}}=\lr{v_n^4}-2\lr{v_n^2}^2\;.
\end{eqnarray}
If harmonic coefficients do not fluctuate event by event, Eq.~\eqref{eq:5} gives $c_n\{2\}_{\mathrm{flow}}=v_n^2$, $c_n\{4\}_{\mathrm{flow}}=-v_n^4$, and $c_n\{4\}_{\mathrm{flow}}$ is expected to be negative. Therefore, the flow coefficients from two- and four-particle cumulants are defined as:
\begin{eqnarray}\label{eq:6}
v_{n}\{2\} =\sqrt{c_n\{2\}}\;,\;\; v_{n}\{4\} =\sqrt[4]{-c_n\{4\}}\;.
\end{eqnarray}

In the standard cumulant method described so far, all $2k$-particle multiplets involved in $\lr{\{2k\}_n}$ are selected using the entire detector acceptance. To further suppress the non-flow correlations that typically involve particles emitted within a localized region in $\eta$, the particles can be grouped into several subevents, each covering a non-overlapping $\eta$ interval~\cite{jjia}. The multi-particle correlations are then constructed by correlating particles between different subevents, further reducing non-flow correlations. This analysis uses the subevent cumulant methods based on two and three subevents as described in the following. 

In the two-subevent cumulant method, the entire event is divided into two subevents, labeled as $a$ and $b$, for example according to $-\eta_{\rm{max}}<\eta_a<0$ and $0<\eta_b<\eta_{\rm{max}}$, where $\eta_{\rm{max}}=2.5$ is the maximum $\eta$ used in the analysis and corresponds to the ATLAS detector acceptance for charged particles. The per-event two- and four-particle azimuthal correlations are then evaluated as:
\begin{eqnarray}\label{eq:7a}
&&\lr{\{2\}_n}_{a|b} = \lr{{\mathrm{e}}^{{\rm i}n(\phi_1^a-\phi_2^b)}} = {\rm{Re}}[{\bm q}_{n,a}{\bm q}^*_{n,b}]\;,\\\label{eq:7b}
&&\lr{\{4\}_n}_{2a|2b} = \lr{{\mathrm{e}}^{{\rm i}n(\phi_1^a+\phi_2^a-\phi_3^b-\phi_4^b)}} = \frac{({\bm q}_{n}^2-\tau_{1}{\bm q}_{2n})_a({\bm q}_{n}^{2}-\tau_{1}{\bm q}_{2n})_b^*}{(1-\tau_{1})_a(1-\tau_{1})_b}\;,
\end{eqnarray}
where the superscript or subscript $a$ ($b$) indicates particles chosen from the subevent $a$ ($b$). Here the four-particle cumulant is defined as:
\begin{eqnarray}\label{eq:8}
c_n^{2a|2b}\{4\}=\llrr{\{4\}_n}_{2a|2b}-2\llrr{\{2\}_n}_{a|b}^2\;.
\end{eqnarray}
The two-subevent method should suppress correlations within a single jet (intra-jet correlations), since each jet usually emits particles into only one subevent.

In the three-subevent cumulant method, the event is divided into three subevents $a$, $b$ and $c$ each covering a unique $\eta$ range, for example $-\eta_{\rm{max}}<\eta_a<-\eta_{\rm{max}}/3$, $|\eta_b|<\eta_{\rm{max}}/3$ and $\eta_{\rm{max}}/3<\eta_c<\eta_{\rm{max}}$. The four-particle azimuthal correlations and cumulants are then evaluated as:
\begin{eqnarray}\label{eq:9}
&&\lr{\{4\}_n}_{2a|b,c}= \lr{{\mathrm{e}}^{{\rm i}n(\phi_1^a+\phi_2^a-\phi_3^b-\phi_4^c)}}= \frac{({\bm q}_{n}^2-\tau_1{\bm q}_{2n})_a{\bm q}_{n,b}^*{\bm q}_{n,c}^*}{(1-\tau_{1})_{a}}\;,\\\label{eq:10}
&&c_n^{2a|b,c}\{4\} \equiv \llrr{\{4\}_n}_{2a|b,c}- 2\llrr{\{2\}_n}_{a|b}\llrr{\{2\}_n}_{a|c} \;,
\end{eqnarray}
where $\llrr{\{2\}_n}_{a|b}$ and $\llrr{\{2\}_n}_{a|c}$ are two-particle correlators defined as in Eq.~\eqref{eq:7a}. Since the two jets in a dijet event usually produce particles in at most two subevents, the three-subevent method further suppresses non-flow contributions from inter-jet correlations associated with dijets. To enhance the statistical precision, the $\eta$ range for subevent $a$ is also interchanged with that for subevent $b$ or $c$, and the resulting three $c_n^{2a|b,c}\{4\}$ values are averaged to obtain the final result.

\section{Detector and trigger}
\label{sec:data}
The ATLAS detector~\cite{Aad:2008zzm} provides nearly full solid-angle coverage around the collision point with tracking detectors, calorimeters, and muon chambers, and is well suited for measurement of multi-particle correlations over a large pseudorapidity range\footnote{ATLAS typically uses a right-handed coordinate system with its origin at the nominal interaction point (IP) in the center of the detector and the $z$-axis along the beam pipe. The $x$-axis points from the IP to the center of the LHC ring, and the $y$-axis points upward. Cylindrical coordinates $(r,\phi)$ are used in the transverse plane, $\phi$ being the azimuthal angle around the beam pipe. By default, the pseudorapidity is defined in terms of the polar angle $\theta$ as $\eta=-\ln\tan(\theta/2)$. However, for asymmetric $\pPb$ or Pb+$p$ collisions, the -$z$ direction is always defined as the direction of the Pb beam.}. The measurements were performed primarily using the inner detector (ID), minimum-bias trigger scintillators (MBTS) and the zero-degree calorimeters (ZDC). The ID detects charged particles within $|\eta| < 2.5$ using a combination of silicon pixel detector, a silicon microstrip detector~(SCT), and a straw-tube transition radiation tracker, all immersed in a 2 T axial magnetic field~\cite{Aad:ID}. An additional pixel layer, the ``insertable B-layer'' (IBL)~\cite{ATLAS-TDR-19} installed between Run 1 (2010--2013) and Run 2 (2015--2018), is available for the Run-2 datasets. The MBTS, rebuilt before Run 2, detects charged particles within $2.1\lesssim|\eta|\lesssim3.9$ using two hodoscopes of counters positioned at $z = \pm$~3.6 m. The ZDC are positioned at $\pm$140~m from the collision point, and detect neutral particles, primarily neutrons and photons, with $|\eta|>8.3$. 

The ATLAS trigger system~\cite{Aad:2012xs} consists of a Level-1 (L1) trigger implemented using a combination of dedicated electronics and programmable logic, and a high-level trigger (HLT) implemented in processors.  The HLT reconstructs charged-particle tracks using methods similar to those applied in the offline analysis, allowing high-multiplicity track (HMT) triggers that select events based on the number of tracks with $\pT>0.4$~GeV associated with the vertex with the largest number of tracks. The different HMT triggers also apply additional requirements on either the transverse energy ($\et$) in the calorimeters or on the number of hits in the MBTS at L1, and on the number of charged-particle tracks reconstructed by the HLT. The $\pp$ and $\pPb$ data were collected using a combination of the minimum-bias and HMT triggers. More details of the triggers used for the $\pp$ and $\pPb$ data can be found in Refs.~\cite{Aad:2016mok,Aaboud:2016yar} and Refs.~\cite{Aad:2014lta,atlas:3}, respectively.

\section{Datasets and Monte Carlo simulations}
\label{sec:datamc}
This analysis uses integrated luminosities of 28~$\mathrm{nb}^{-1}$ of $\pPb$ data recorded at $\sqrtsnn = 5.02$~TeV, 0.17~$\mathrm{pb}^{-1}$ of $\pp$ data recorded at $\sqrt{s} = 5.02$~TeV, and 0.9 $\mathrm{pb}^{-1}$ of $\pp$ data recorded at $\sqrt{s} = 13$~TeV, all taken by the ATLAS experiment at the LHC. The $\pPb$ data were mainly collected in 2013, but also include 0.3~$\mathrm{nb}^{-1}$ data collected in November 2016 which increases the number of events at moderate multiplicity (see Section~\ref{sec:sel}). During both $\pPb$ runs, the LHC was configured with a 4~TeV proton beam and a 1.57~TeV per-nucleon Pb beam that together produced collisions at $\sqrtsnn = 5.02$~TeV, with a rapidity shift of $0.465$ of the nucleon--nucleon center-of-mass frame towards the proton beam direction relative to the ATLAS rest frame. The direction of the Pb beam is always defined to have negative pseudorapidity. The 5.02~TeV $\pp$ data were collected in November 2015. The 13~TeV $\pp$ data were collected during several special low-luminosity runs of the LHC in 2015 and 2016.

Monte Carlo (MC) simulated event samples are used to determine the track reconstruction efficiency (Section~\ref{sec:sel}). The 13 TeV and 5.02 TeV $\pp$ data were simulated by the $\PYTHIA{8}$ MC event generator~\cite{Sjostrand:2007gs} using the A2 set of tuned parameters with MSTW2008LO parton distribution functions~\cite{atlas:6}. The HIJING event generator~\cite{Gyulassy:1994ew} was used to produce p+Pb collisions with the same energy and the same boost of the center-of-mass system as in the data. The detector response was simulated using $\GEANT{4}$~\cite{Agostinelli:2002hh,Aad:2010ah} with detector conditions matching those during the data-taking. The simulated events and data events are reconstructed with the same algorithms, including those for track reconstruction.

\section{Event and track selection}
\label{sec:sel}
The offline event selection for the $\pPb$ and $\pp$ data requires at least one reconstructed vertex with its longitudinal position satisfying $|\zvtx|< 100$~mm. The vertex is required to have at least two associated tracks with $\pT>0.4$~GeV. The mean collision rate per bunch crossing $\mu$ was approximately 0.03 for the 2013 $\pPb$ data, 0.001--0.006 for the 2016 $\pPb$ data, 0.02--1.5 for 5.02~TeV $\pp$ data and 0.002--0.8 for the 13~TeV $\pp$ data. In order to suppress additional interactions in the same bunch crossing (referred to as pileup) in $\pp$ collisions, events containing additional vertices with at least four associated tracks are rejected. In $p$+Pb collisions, events with more than one good vertex, defined as any vertex for which the scalar sum of the $\pT$ of the associated tracks is greater than 5 GeV, are rejected. The remaining pileup events are further suppressed by using the signal in the ZDC on the Pb-fragmentation side. This signal is calibrated to the number of detected neutrons ($N_{n}$) by using the location of the peak corresponding to a single neutron. The distribution of $N_{n}$ in events with pileup is broader than that for the events without pileup. Hence a simple requirement on the ZDC signal distribution is used to further suppress events with pileup, while retaining more than 98\% of the events without pileup. The impact of residual pileup, at a level of $\lesssim 10^{-3}$, is studied by comparing the results obtained from data with different $\mu$ values.

Charged-particle tracks and collision vertices are reconstructed using the same algorithms and methods applied in previous minimum-bias $\pp$ and $\pPb$ measurements~\cite{Aad:2014lta,Aad:2016mok,Aad:2015gqa}. For the 2013 $p$+Pb analysis, tracks are required to have a $\pT$-dependent minimum number of hits in the SCT. The transverse ($d_0$) and longitudinal ($z_0$ $\sin\theta$) impact parameters of the track relative to the primary vertex are both required to be less than 1.5 mm. A more detailed description of the track selection for the 2013 $\pPb$ data can be found in Ref.~\cite{Aad:2014lta}.

For all the data taken since the start of Run 2, the track selection criteria make use of the IBL, as described in Refs.~\cite{Aad:2016mok,Aad:2015gqa}. Furthermore, the requirements of $|d_0^{\rm{BL}}|<1.5$ mm and $|z_0\sin\theta|<1.5$~mm are applied, where $d_0^{\rm{BL}}$ is the transverse impact parameter of the track relative to the beam line (BL).

The cumulants are calculated using tracks passing the above selection requirements, and having $|\eta|<2.5$ and $0.3<\pT <3$~GeV or $0.5<\pT <5$~GeV. These two $\pT$ ranges are chosen because they were often used in previous ridge measurements at the LHC~\cite{Khachatryan:2015waa,Khachatryan:2016txc,Aad:2015gqa,Aaboud:2016yar,atlas:4}. However, to count the number of reconstructed charged particles for event-class definition (denoted by $\nchrec$), tracks with $\pT>0.4$~GeV and $|\eta|<2.5$ are used for compatibility with the requirements in the HLT selections described above. Due to different trigger requirements,  most of the $\pPb$ events with $\nchrec>150$ are provided by the 2013 dataset, while the 2016 dataset provides most of the events at lower $\nchrec$.

The efficiency of the combined track reconstruction and selection requirements in data is estimated using the MC samples reconstructed with the same tracking algorithms and the same track selection requirements. Efficiencies, $\epsilon(\eta,\pT)$, are evaluated as a function of track $\eta$, $\pT$ and the number of reconstructed charged-particle tracks, but averaged over the full range in azimuth. For all collision systems, the efficiency increases by about 4\% as $\pT$ increases from 0.3 GeV to 0.6 GeV. Above 0.6 GeV, the efficiency is independent of $\pT$ and reaches 86\% (72\%) at $\eta\approx 0$ ($|\eta| > 2$) for $pp$ collisions  and 83\% (70\%) for $\pPb$ collisions, respectively. The efficiency is independent of the event multiplicity for $\nchrec > 40$. For lower-multiplicity events the efficiency is smaller by up to a few percent due to broader $d_0^{\rm{BL}}$ and $z_0\sin\theta$ distributions.

The rate of falsely reconstructed charged-particle tracks is also estimated and found to be negligibly small in all datasets. This rate decreases with increasing $\pT$, and even at the lowest transverse momenta of 0.2 GeV it is below 1\% of the total number of tracks. Therefore, there is no correction for the presence of these tracks in the analysis. 

In the simulated events, the reconstruction efficiency reduces the measured charged-particle multiplicity relative to the generated multiplicity for primary charged particles. The multiplicity correction factor $b$ is used to correct $\nchrec$ to obtain the efficiency-corrected number of charged particles per event, $\nch=b\lr{\nchrec}$. The value of the correction factor is found to be independent of $\nchrec$ in the range used in this analysis. Its value and the associated uncertainties are $b=1.29$ $\pm$ 0.05 for the 2013 $\pPb$ collisions and $b=1.18\pm$~0.05 for Run-2 $\pPb$ and $\pp$ collisions~\cite{Aaboud:2016jnr}. Both $c_n\{4\}$ and $v_n\{4\}$ are then studied as a function of $\nch$.
\section{Data analysis}
\label{sec:corr}
The multi-particle cumulants are calculated in three steps using charged particles with $|\eta|<2.5$. In the first step, the multi-particle correlators $\lr{\{2k\}_n}$ from Eqs.~\eqref{eq:1},~\eqref{eq:2}, \eqref{eq:7a}, \eqref{eq:7b} and \eqref{eq:9} are calculated for each event from particles in one of two $\pT$ ranges, $0.3<\pT<3$ GeV and $0.5<\pT<5$ GeV. 

In the second step, the correlators $\lr{\{2k\}_n}$ are averaged over events with the same $\nchb$, the number of reconstructed charged particles in a given $\pT$ range, to obtain $\llrr{\{2k\}_n}$ and $c_n\{2k\}$ from Eqs.~\eqref{eq:4},\eqref{eq:8} and \eqref{eq:10}. In a previous study~\cite{atlas:4}, it was observed that the $c_n\{2k\}$ values varied with the exact definition of $\nchb$. This is because different definitions of $\nchb$ lead to different multiplicity fluctuations and therefore different non-flow correlations associated with these multiplicity fluctuations. The observed dependence of $c_n\{2k\}$ on the definition of $\nchb$ has been attributed to the change in the non-flow correlations when $\nchb$ is changed~\cite{atlas:4}.

In order to further test the sensitivity of $c_n\{2k\}$ to the exact definition of $\nchb$, four different $\pT$ requirements are used to define $\nchb$ as follows:  when $\lr{\{2k\}_n}$ is calculated in the range $0.3<\pT<3$ GeV, $\nchb$ is evaluated in four different track $\pT$ ranges: $0.3<\pT<3$ GeV, $\pT>0.2$ GeV, $\pT>0.4$ GeV and $\pT>0.6$ GeV.  When $\lr{\{2k\}_n}$ is calculated in $0.5<\pT<5$ GeV, $\nchb$ is evaluated in four different track $\pT$ ranges: $0.5<\pT<5$ GeV, $\pT>0.2$ GeV, $\pT>0.4$ GeV and $\pT>0.6$ GeV. In each case, the $c_n\{2k\}$ value is first calculated for events with the same $\nchb$; the $c_n\{2k\}$ values are then combined in the broader $\nchb$ range of the event ensemble to obtain statistically significant results. 

In the third step, the $c_n\{2k\}$ and $v_n\{2k\}$ values obtained for a given $\nchb$ are mapped to a given $\lr{\nchc}$, the average number of reconstructed charged particles with $\pT>0.4$ GeV. The mapping procedure is necessary so that $c_n\{2k\}$ obtained for different $\nchb$ can be compared using a common $x$-axis defined by $\lr{\nchc}$. The $\lr{\nchc}$ value is then converted to $\nch$, the efficiency-corrected average number of charged particles with $\pT > 0.4$ GeV, as discussed in Section~\ref{sec:sel}.

In order to account for detector inefficiencies and non-uniformity, particle weights used in Eq.~\eqref{eq:3} are defined as:
\begin{eqnarray}\label{eq:11}
w_i(\phi,\eta,\pT) = d(\phi,\eta)/\epsilon(\eta,\pT)\;.
\end{eqnarray}
The additional weight factor $d(\phi,\eta)$ accounts for non-uniformities in the azimuthal acceptance of the detector as a function of $\eta$. All reconstructed charged particles with $\pT>0.2$ GeV are entered into a two-dimensional histogram $N(\phi,\eta)$, and the weight factor is then obtained as $d(\phi,\eta) \equiv \lr{N(\eta)}/N(\phi,\eta)$, where $\lr{N(\eta)}$ is the track density averaged over $\phi$ in the given $\eta$ bin. This procedure removes most $\phi$-dependent non-uniformity from track reconstruction for any azimuthal correlation analysis~\cite{atlas:4}.

\section{Systematic uncertainties}
\label{sec:syscheck}
The main sources of systematic uncertainty are related to the detector azimuthal non-uniformity, track selection, track reconstruction efficiency, trigger efficiency and pileup. Most of the systematic uncertainties enter the analysis through the particle weights, Eq.~\eqref{eq:11}. Since $c_2\{4\}$ often changes sign in the low $\nch$ region, the absolute uncertainties (instead of relative uncertainties) in $c_2\{4\}$ are determined for each source. The uncertainties are typically of the order of $10^{-6}$, which translates into an absolute uncertainty of $\sqrt[4]{10^{-6}}= 0.032$ for zero flow signal.

The effect of detector azimuthal non-uniformity is accounted for using the weight factor $d(\phi,\eta)$. The impact of the reweighting procedure is studied by fixing the weight to unity and repeating the analysis. The results are mostly consistent with the nominal results within statistical uncertainties. As a cross-check, the multi-particle correlations are calculated using a mixed-event procedure, where each particle in a $2k$-multiplet is selected from a different event with similar $\nchrec$ ($|\Delta\nchrec|<10$) and similar $\zvtx$ ($|\Delta\zvtx|<10$~mm). The particle weights defined in Eq.~\eqref{eq:11} are applied for each particle forming the mixed event. The $c_2\{4\}$ signal obtained from the mixed events is less than $0.2\times 10^{-6}$ in all datasets.

The systematic uncertainty associated with the track selection is estimated by tightening the $|d_0|$ and $|z_0\sin\theta|$ requirements. For each variation, the tracking efficiency is re-evaluated and the analysis is repeated. The maximum differences from the nominal results are observed to be less than $0.3\times 10^{-6}$, $0.2\times 10^{-6}$ and $0.1\times 10^{-6}$ in 5.02~TeV $\pp$, 13~TeV $\pp$ and $\pPb$ collisions, respectively. 

Previous measurements indicate that the azimuthal correlations (both the flow and non-flow components) have a strong dependence on $\pT$, but a relatively weak dependence on $\eta$~\cite{Aad:2014lta,Aaboud:2016yar}. Therefore, $\pT$-dependent systematic effects in the track reconstruction efficiency could affect $c_n\{2k\}$ and $v_n\{2k\}$ values. The uncertainty in the track reconstruction efficiency is mainly due to differences in the detector conditions and material description between the simulation and the data. The efficiency uncertainty varies between 1\% and 4\%, depending on track $\eta$ and $\pT$~\cite{Aaboud:2016yar,atlas:4}. Its impact on multi-particle cumulants is evaluated by repeating the analysis with the tracking efficiency varied up and down by its corresponding uncertainty as a function of $\pT$. For the standard cumulant method, which is more sensitive to jets and dijets, the evaluated uncertainty amounts to (0.1--1.5)$\times 10^{-6}$ in $\pp$ collisions and less than $0.3\times 10^{-6}$ in $\pPb$ collisions for $\nch>50$. For the two- and three-subevent methods, the evaluated uncertainty is typically less than $0.3\times 10^{-6}$ for most of the $\nch$ ranges.

Most events used in the analysis are collected with the HMT triggers with several $\nchrec$ thresholds. In order to estimate the possible bias due to trigger inefficiency as a function of $\nch$, the offline $\nchrec$ requirements are changed such that the HMT trigger efficiency is at least 50\% or 80\%. The results are obtained independently for each variation. These results are found to be consistent with each other for the two- and three-subevent methods, and show a small difference for the standard cumulant method in the low $\nch$ region. The nominal analysis is performed using the 50\% efficiency selection and the differences between the nominal results and those from the 80\% efficiency selection are used as a systematic uncertainty. The change amounts to (0.1--0.7)$\times 10^{-6}$. 

In this analysis, a pileup rejection criterion is applied to reject events containing additional vertices. In order to check the impact of residual pileup, the analysis is repeated without the pileup rejection criterion, and no difference is observed. For the 5.02 and 13~TeV $\pp$ datasets, which have relatively high pileup, the data is divided into two samples based on the $\mu$ value: $\mu>0.4$ and $\mu<0.4$, and the results are compared. The average $\mu$ values differ by a factor of two between the two samples, and the difference in $c_2\{4\}$ is found to be less than 0.5$\times 10^{-6}$.

To check the impact of dijet events, where both jets have pseudorapidities close to the boundaries of relevant subevent regions, the three-subevent cumulants are calculated by requiring a $\Delta\eta=0.5$ gap between the adjacent regions. The results are found to be consistent with the nominal result. 

The systematic uncertainties from different sources are added in quadrature to determine the total systematic uncertainty. The uncertainty is (0.1--1)$\times 10^{-6}$ for two- and three-subevent methods in the region $\nch>50$, where there is a negative $c_2\{4\}$ signal. The total systematic uncertainty for the standard method is typically about a factor of two larger.

The systematic uncertainty studies described above are also carried out for $c_3\{4\}$, and the absolute uncertainties are found to be smaller than those for $c_2\{4\}$, presumably because $c_3\{4\}$ is less sensitive to the influence from dijets.

\vspace*{-0.3cm}
\section{Results}
\label{sec:result}
\vspace*{-0.2cm}
\subsection{Dependence on the event-class definition}
This section presents the sensitivity of $c_2\{4\}$ to $\nchb$, which defines the event class used to calculate $\llrr{\{2\}_n}$ and $\llrr{\{4\}_n}$ in Eqs.~\eqref{eq:8}--\eqref{eq:10}. The discussion is based on results obtained from the 13~TeV $\pp$ data, but the observations for the 5.02~TeV $\pp$ and $\pPb$ data are qualitatively similar.

\begin{figure}[h!]
\begin{center}
\includegraphics[width=0.5\linewidth]{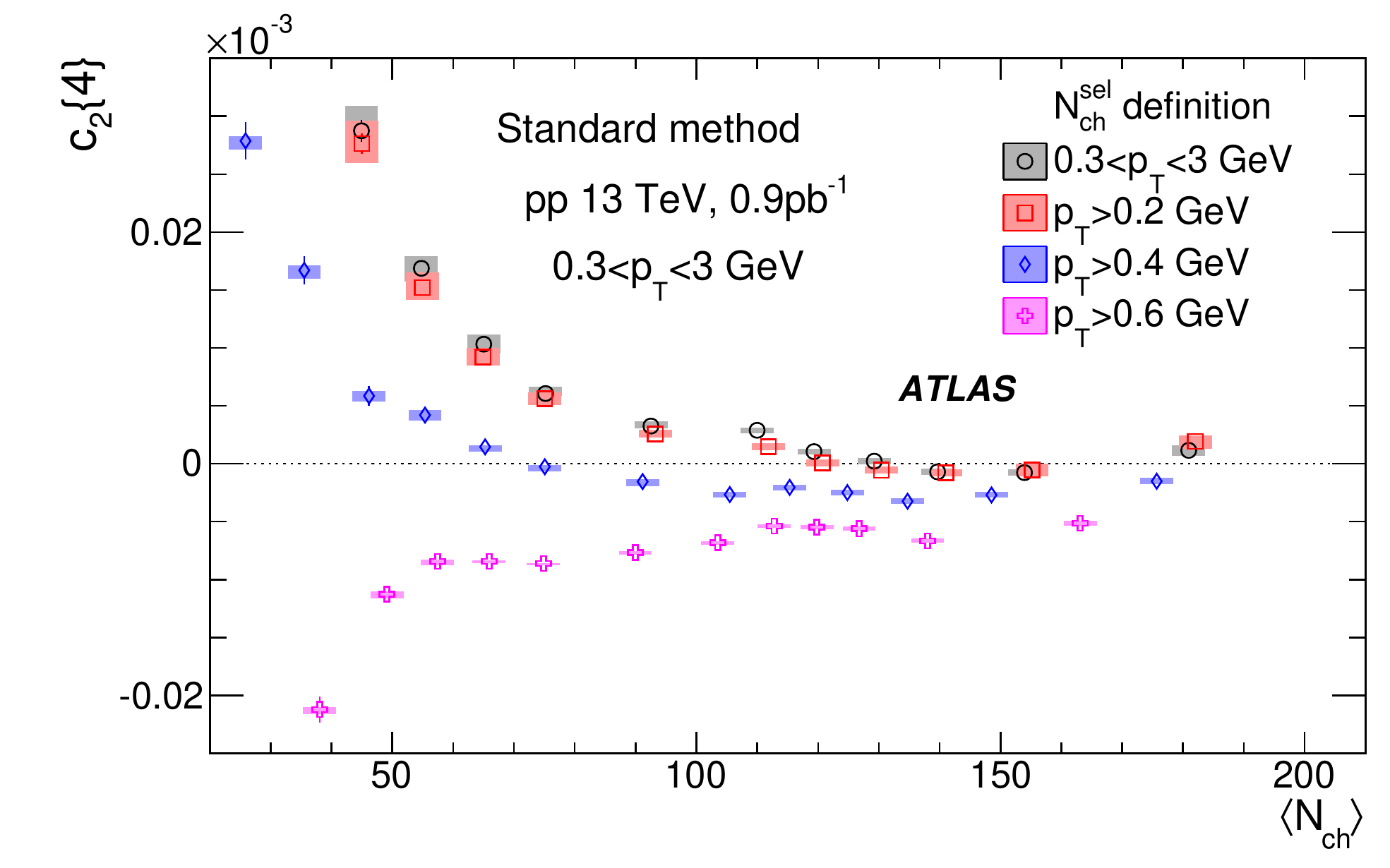}\includegraphics[width=0.5\linewidth]{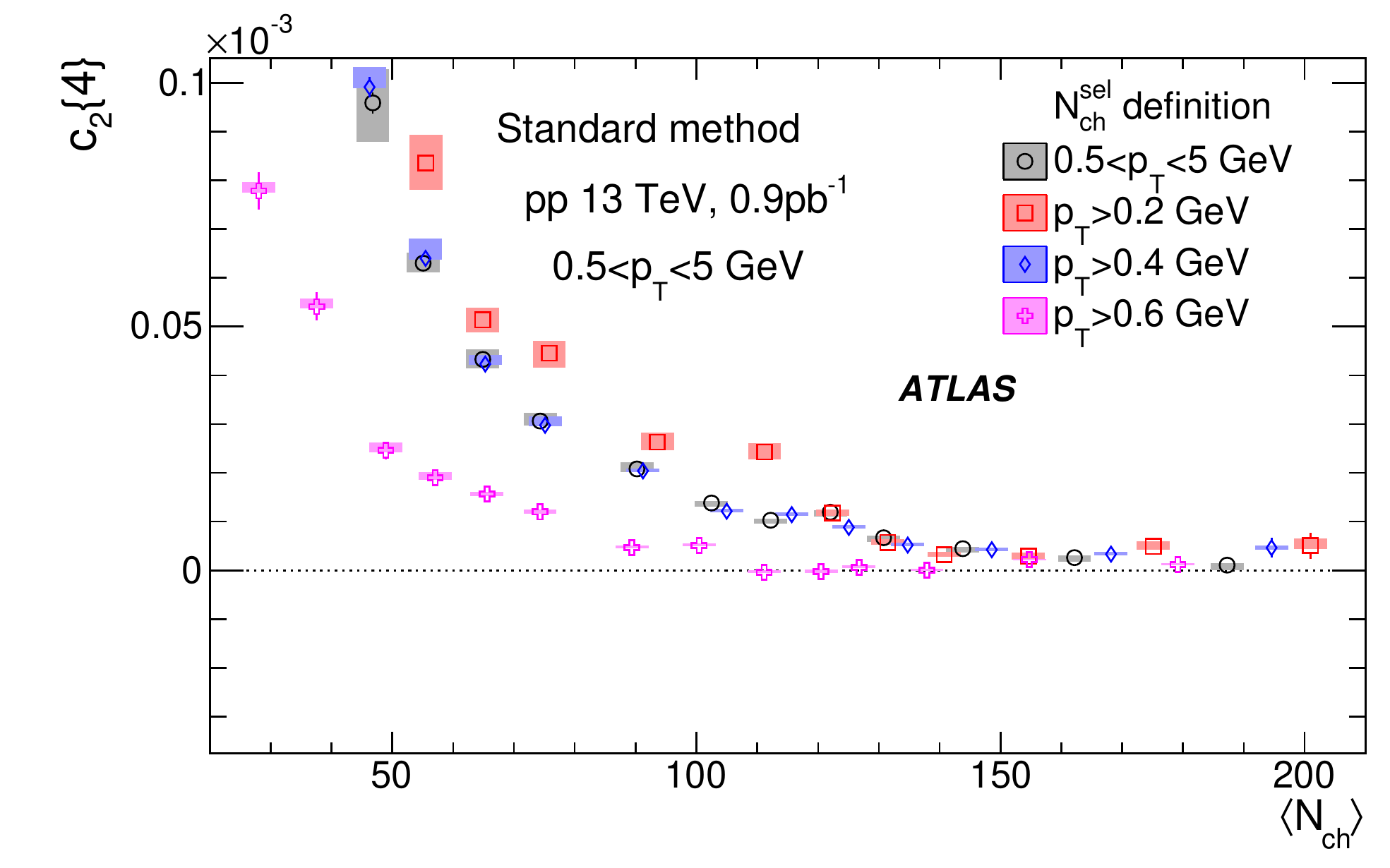}
\end{center}\vspace*{-0.7cm}
\caption{\label{fig:1} The $c_2\{4\}$ values calculated for charged particles with $0.3<\pT<3$~GeV (left panel) and $0.5<\pT<5$~GeV  (right panel) with the standard cumulant method from the 13~TeV $\pp$ data. The event averaging is performed for $\nchb$ calculated for various $\pT$ selections as indicated in the figure, which is then mapped to $\nch$, the average number of charged particles with $\pT>0.4$~GeV. The error bars and shaded boxes represent the statistical and systematic uncertainties, respectively.}
\end{figure}

Figure~\ref{fig:1} shows the $c_2\{4\}$ values obtained using the standard method for four event-class definitions based on $\nchb$. The $c_2\{4\}$ values changes dramatically as the event-class definition is varied, which, as points out in Ref.~\cite{jjia}, reflects different amount of non-flow flucuations associated with different $\nchb$. The $c_2\{4\}$ values for $0.3<\pT<3$~GeV become negative when the reference $\nchb$ is obtained for $\pT>0.4$~GeV or higher, but the four cases do not converge to the same $c_2\{4\}$ values. On the other hand, $c_2\{4\}$ values for $0.5<\pT<5$~GeV are always positive, independent of the definition of $\nchb$. These behaviors suggest that the $c_2\{4\}$ values from the standard method are strongly influenced by non-flow effects in all $\nch$ and $\pT$ ranges. Therefore the previously observed negative $c_2\{4\}$ in $pp$ collisions for $0.3<\pT<3$~GeV and $\nchb$ with $\pT>0.4$~GeV~\cite{Khachatryan:2016txc} may be dominated by non-flow correlations instead of long-range collective flow.

\begin{figure}[h!]
\begin{center}
\includegraphics[width=0.5\linewidth]{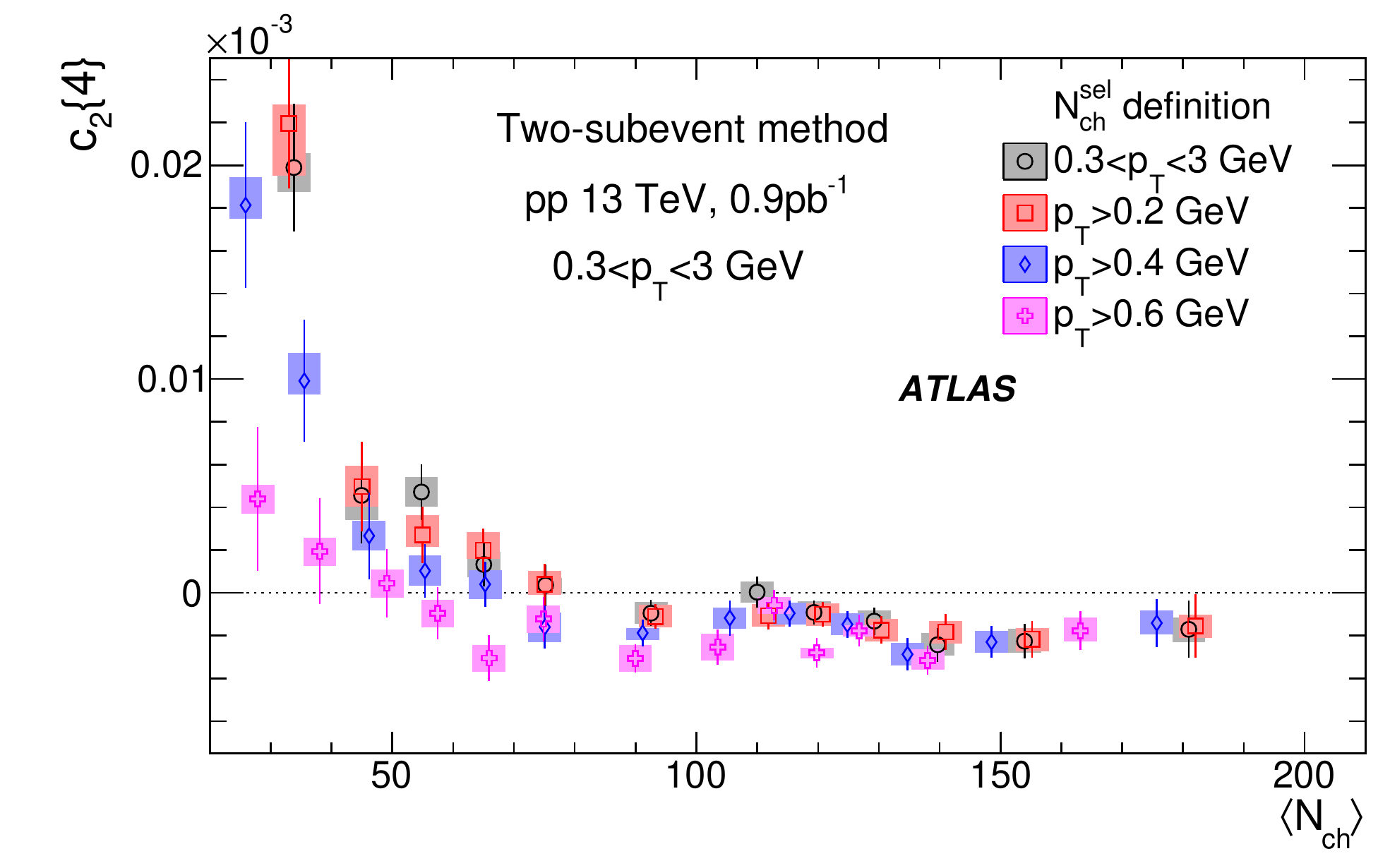}\includegraphics[width=0.5\linewidth]{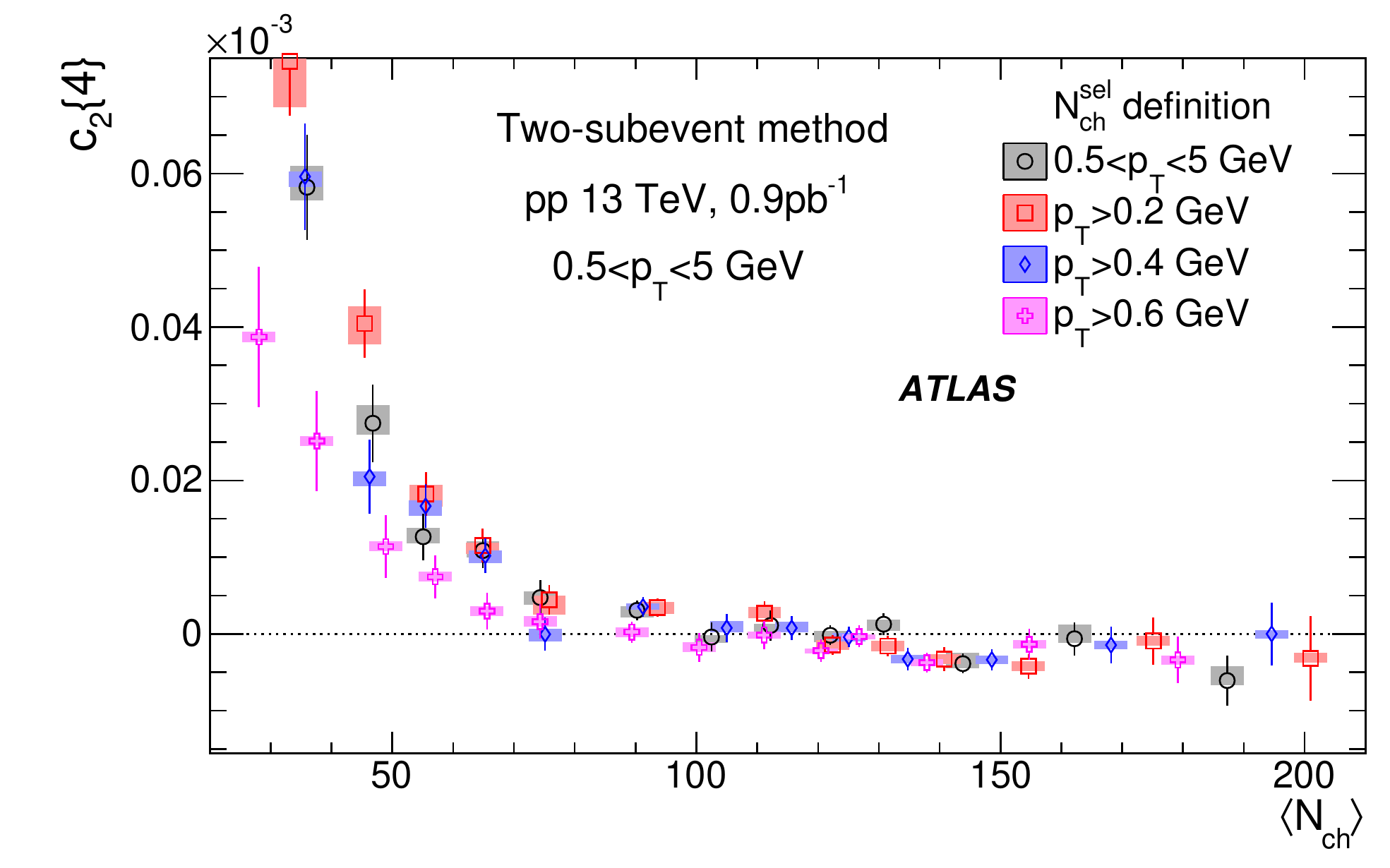}
\end{center}
\caption{\label{fig:2} The $c_2\{4\}$ values calculated for charged particles with $0.3<\pT<3$~GeV (left panel) and $0.5<\pT<5$~GeV (right panel) with the two-subevent cumulant method from the 13~TeV $\pp$ data. The event averaging is performed for $\nchb$ calculated for various $\pT$ selections as indicated in the figure, which is then mapped to $\nch$, the average number of charged particles with $\pT>0.4$~GeV. The error bars and shaded boxes represent the statistical and systematic uncertainties, respectively.}
\end{figure}

Figure~\ref{fig:2} shows that the $c_2\{4\}$ values calculated using the two-subevent method are closer to each other among different event-class definitions.
 The $c_2\{4\}$ values decrease gradually with $\nch$ and become negative for $\nch>70$ when $c_2\{4\}$ is calculated in the range $0.3<\pT<3$~GeV range and for $\nch>150$ when $c_2\{4\}$ is calculated in the range $0.5<\pT<5$~GeV. Therefore, the $c_2\{4\}$ values from the two-subevent method are more sensitive to long-range ridge correlations, but nevertheless may still be affected by non-flow effects, especially in the low $\nch$ region and higher $\pT$.

\begin{figure}[h!]
\begin{center}
\includegraphics[width=0.5\linewidth]{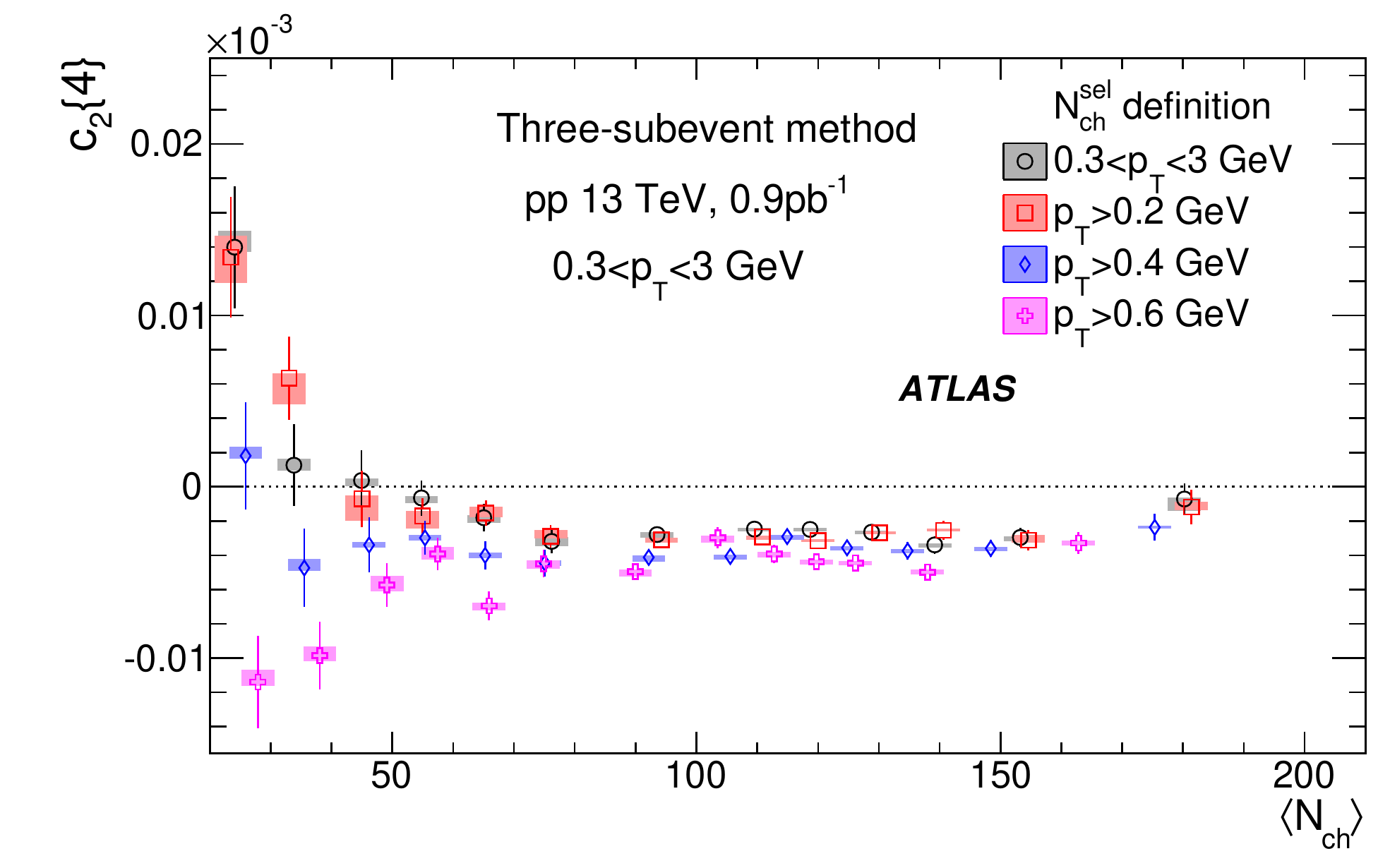}\includegraphics[width=0.5\linewidth]{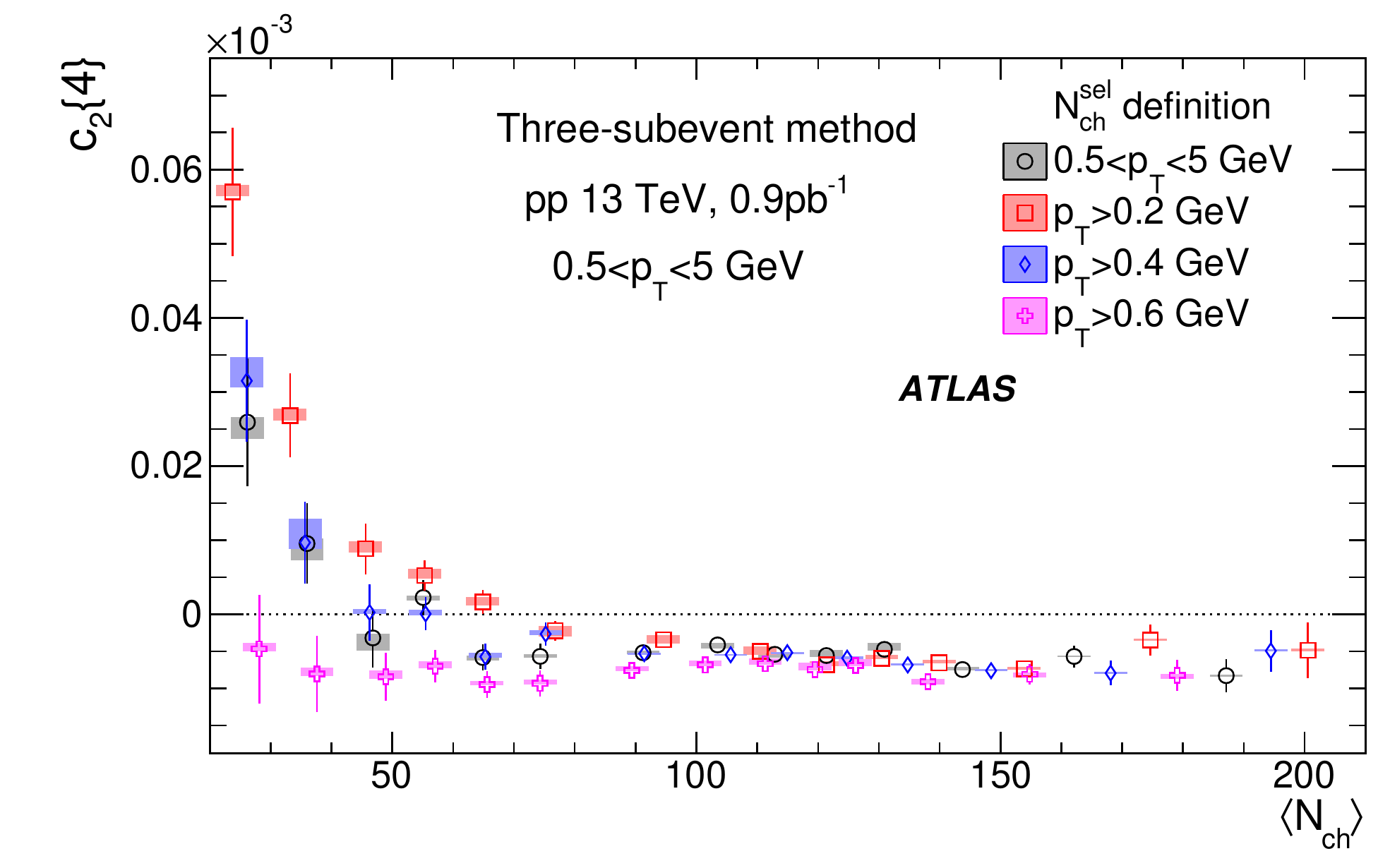}
\end{center}
\caption{\label{fig:3} The $c_2\{4\}$ values calculated for charged particles with $0.3<\pT<3$~GeV (left panel) and $0.5<\pT<5$~GeV (right panel) with the three-subevent cumulant method from the 13~TeV $\pp$ data. The event averaging is performed for $\nchb$ calculated for various $\pT$ selections as indicated in the figure, which is then mapped to $\nch$, the average number of charged particles with $\pT>0.4$~GeV. The error bars and shaded boxes represent the statistical and systematic uncertainties, respectively.}
\end{figure}

Figure~\ref{fig:3} shows the results from the three-subevent method. For most of the $\nch$ range, the $c_2\{4\}$ values are negative, i.e., having the sign expected for long-range ridge correlations. The $c_2\{4\}$ values show some sensitivity to the definition of the reference $\nchb$ but they are close to each other for all definitions in the region $\nch>100$. This suggests that the residual non-flow effects may still be important at small $\nch$, but are negligible at $\nch>100$. It is also observed that the $c_2\{4\}$ values for $0.5<\pT<5$~GeV are more negative than those for $0.3<\pT<3$~GeV, which is consistent with the observation that the $v_2$ value associated with the long-range collectivity increases with $\pT$~\cite{Aad:2014lta,Aaboud:2016yar}. 

Given the relatively small dependence of $c_2\{4\}$ on the reference $\nchb$ in the three-subevent method, the remaining discussion focuses on cases where the reference $\nchb$ is calculated in the same $\pT$ ranges as those used for calculating $c_2\{4\}$, i.e.\ $0.3<\pT<3$~GeV and $0.5<\pT<5$~GeV.

\subsection{Comparison between different cumulant methods}
\begin{figure}[!h]
\begin{center}
\includegraphics[width=0.5\linewidth]{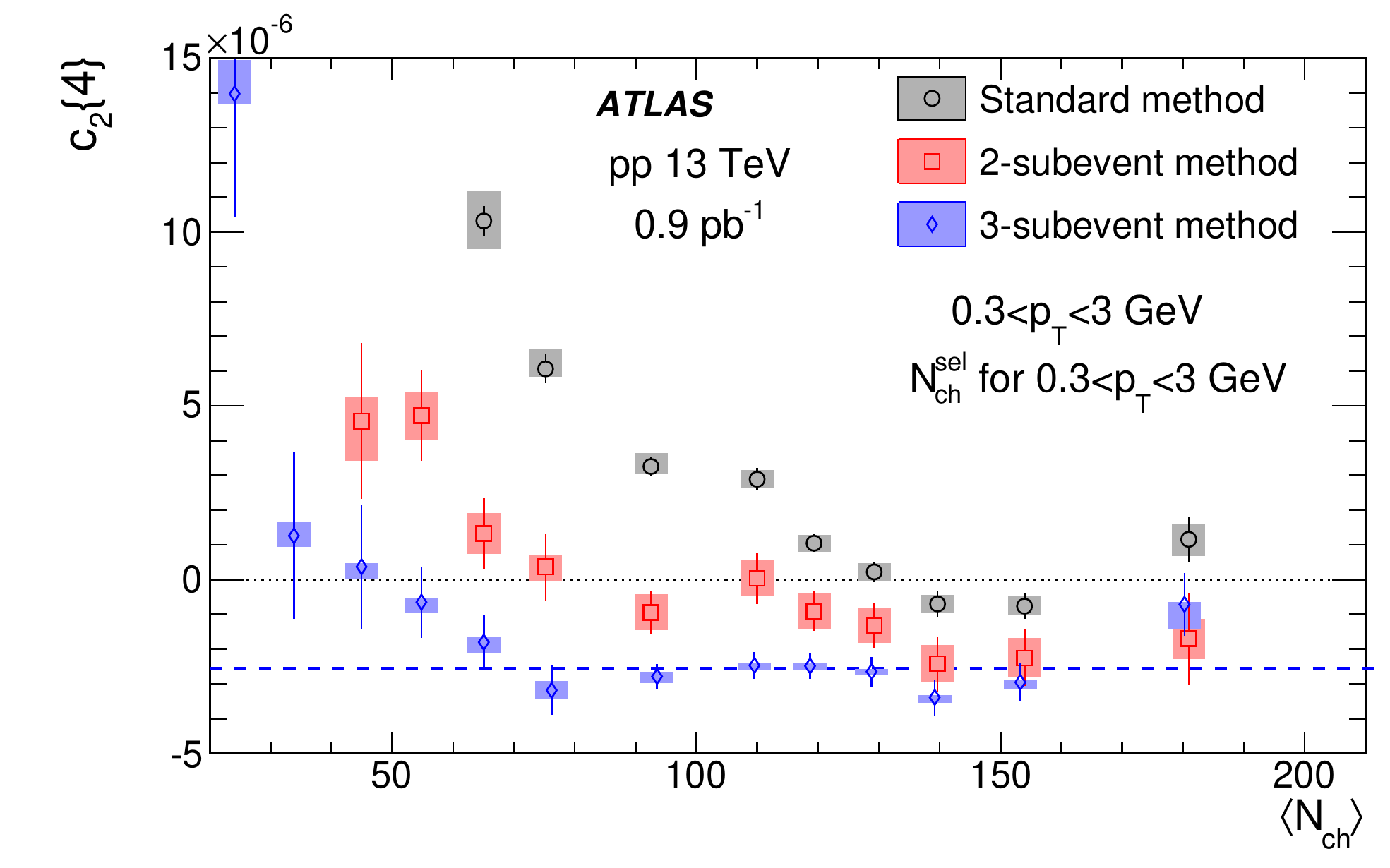}\includegraphics[width=0.5\linewidth]{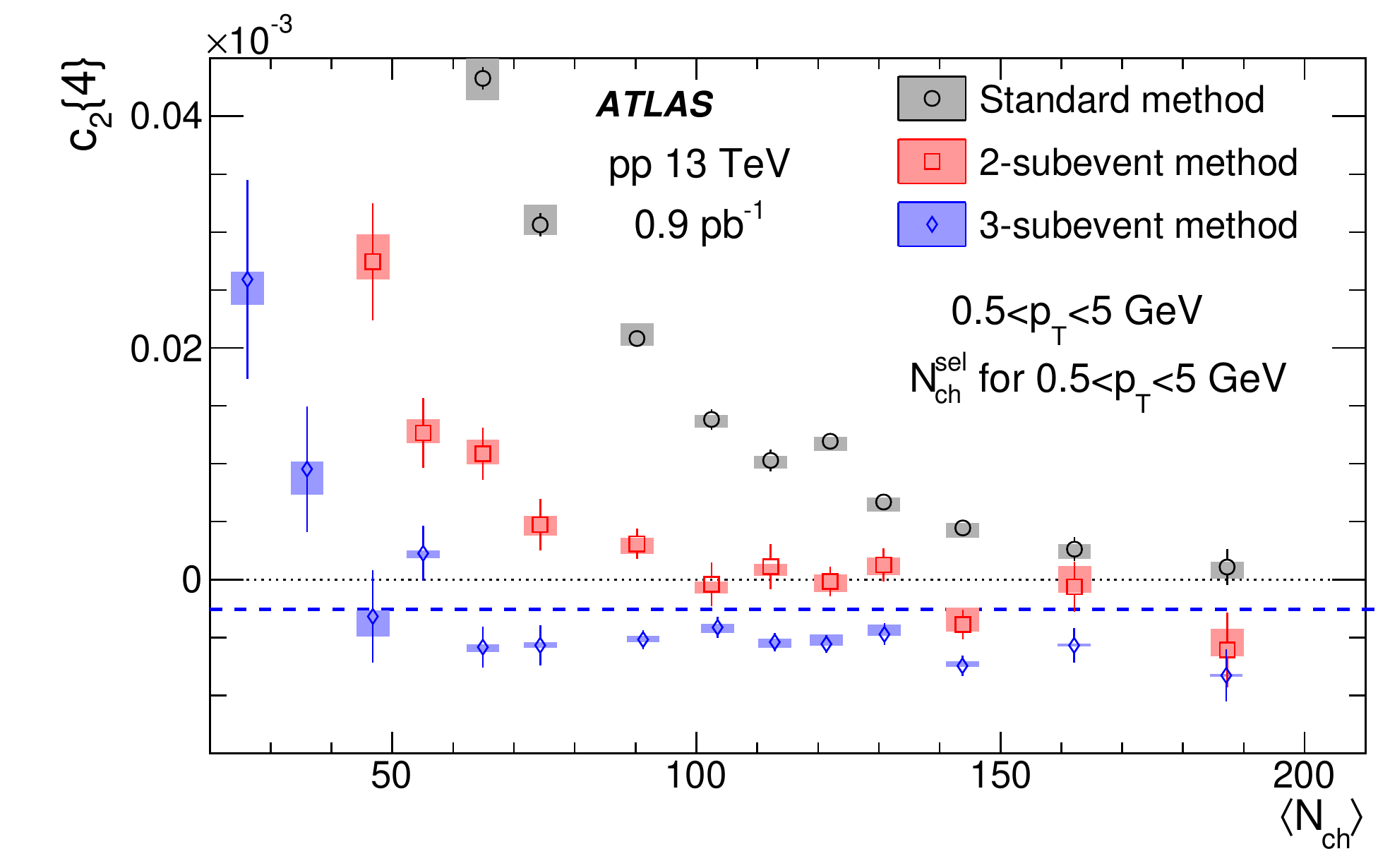}
\end{center}\vspace*{-0.6cm}
\caption{\label{fig:4} The $c_2\{4\}$ values calculated for charged particles with $0.3<\pT<3$~GeV (left panel) and $0.5<\pT<5$~GeV (right panel) compared for the three cumulant methods from the 13~TeV $\pp$ data. The event averaging is performed for $\nchb$ calculated for the same $\pT$ range, which is then mapped to $\nch$, the average number of charged particles with $\pT>0.4$~GeV. The dashed line indicates the $c_2\{4\}$ value corresponding to a 4\% $v_2$ signal. The error bars and shaded boxes represent the statistical and systematic uncertainties, respectively.}
\end{figure}
\begin{figure}[!h]
\begin{center}
\includegraphics[width=0.5\linewidth]{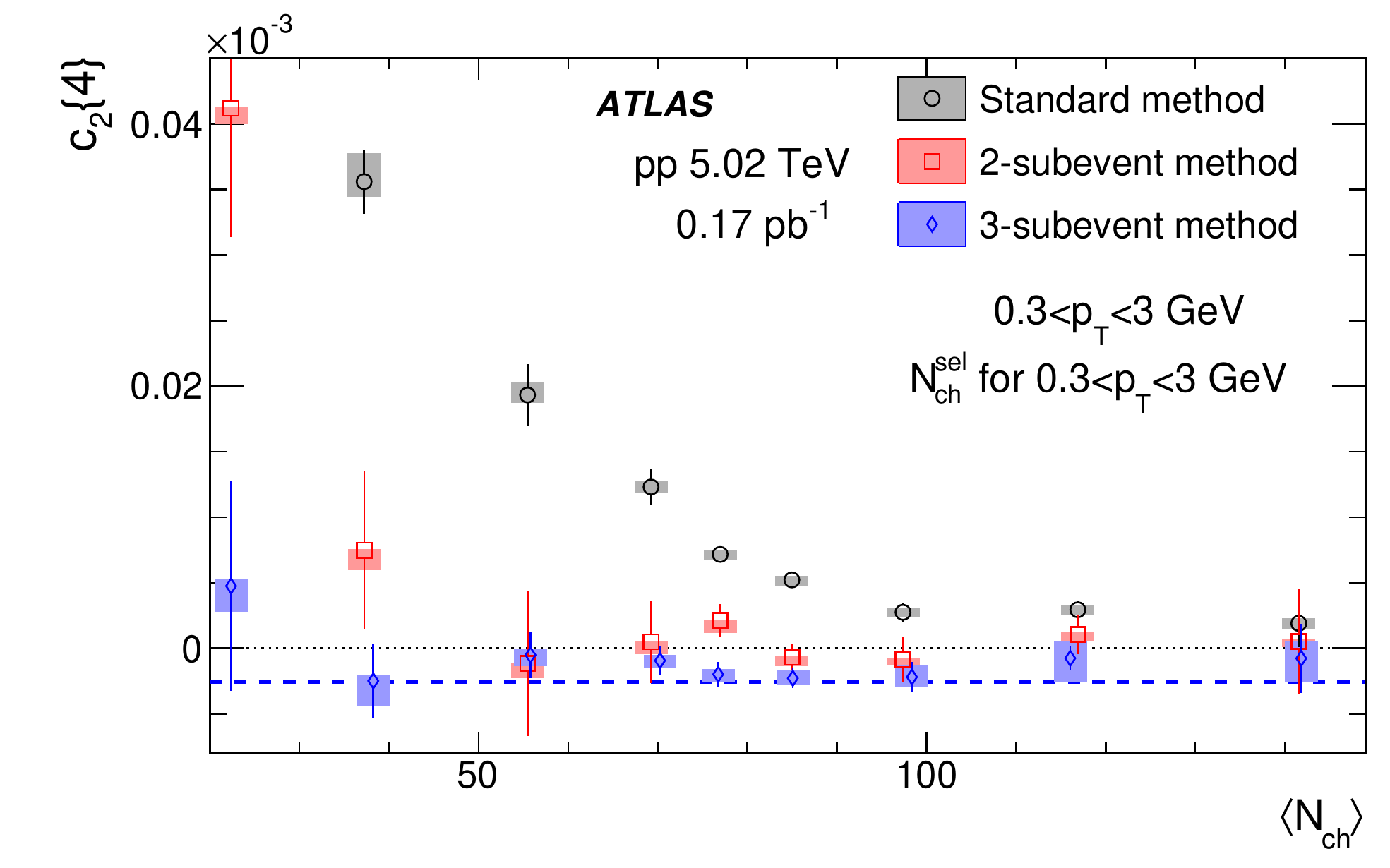}\includegraphics[width=0.5\linewidth]{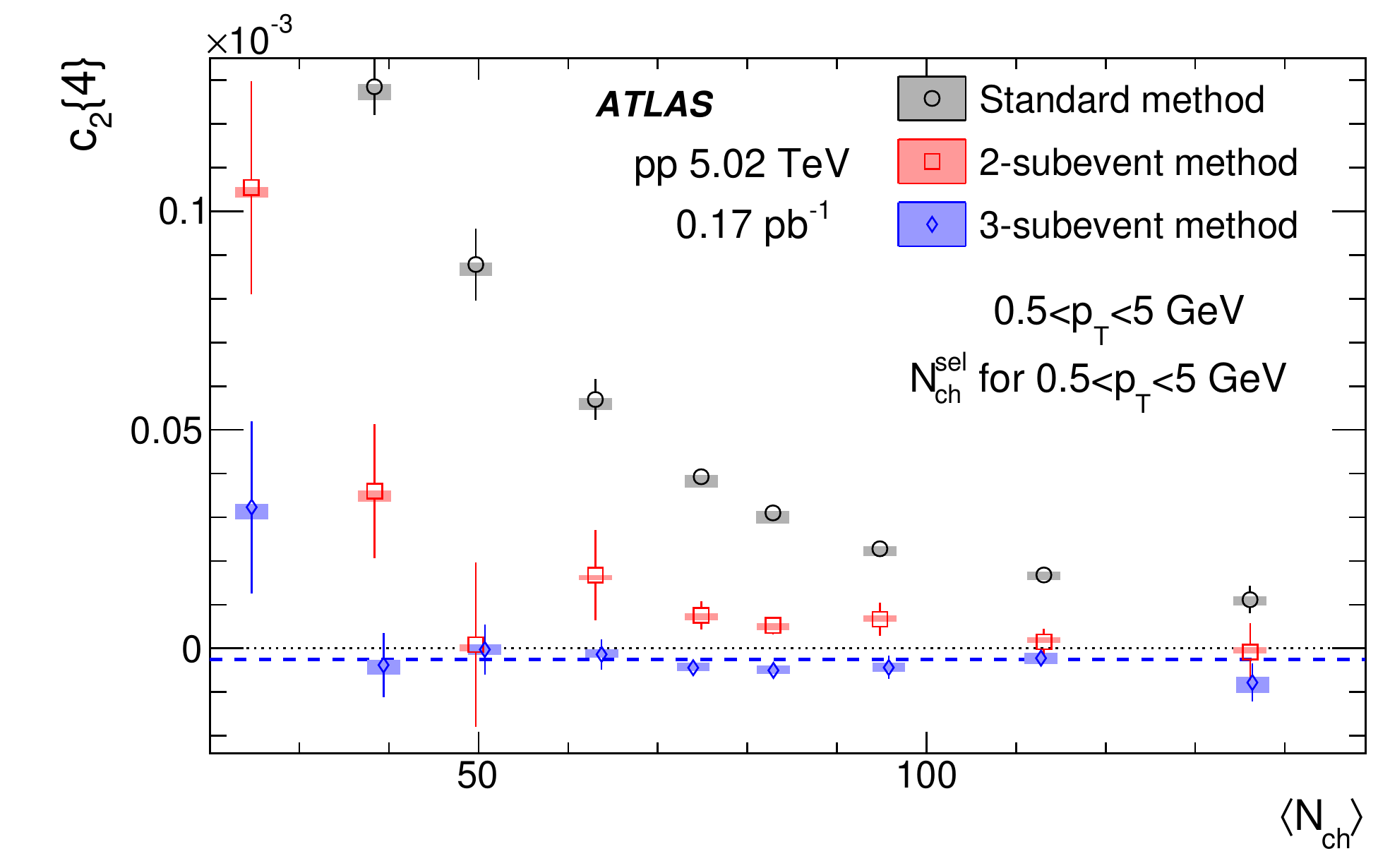}
\end{center}\vspace*{-0.6cm}
\caption{\label{fig:5} The $c_2\{4\}$ values calculated for charged particles with $0.3<\pT<3$~GeV (left panel) and $0.5<\pT<5$~GeV (right panel) compared for the three cumulant methods from the 5.02~TeV $\pp$ data. The event averaging is performed for $\nchb$ calculated for the same $\pT$ range, which is then mapped to $\nch$, the average number of charged particles with $\pT>0.4$~GeV. The dashed line indicates the $c_2\{4\}$ value corresponding to a 4\% $v_2$ signal. The error bars and shaded boxes represent the statistical and systematic uncertainties, respectively.}
\end{figure}
\begin{figure}[!h]
\begin{center}
\includegraphics[width=0.5\linewidth]{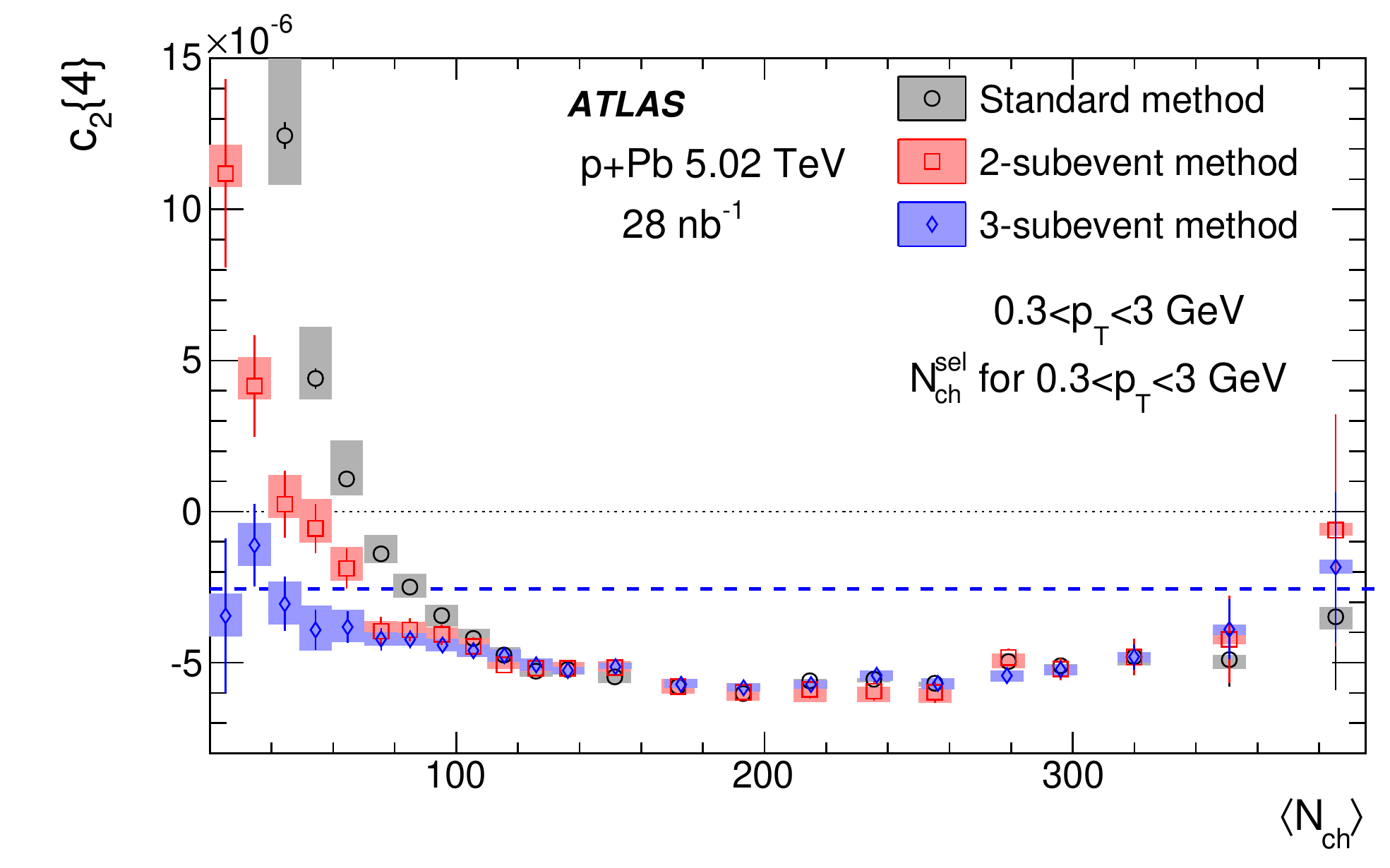}\includegraphics[width=0.5\linewidth]{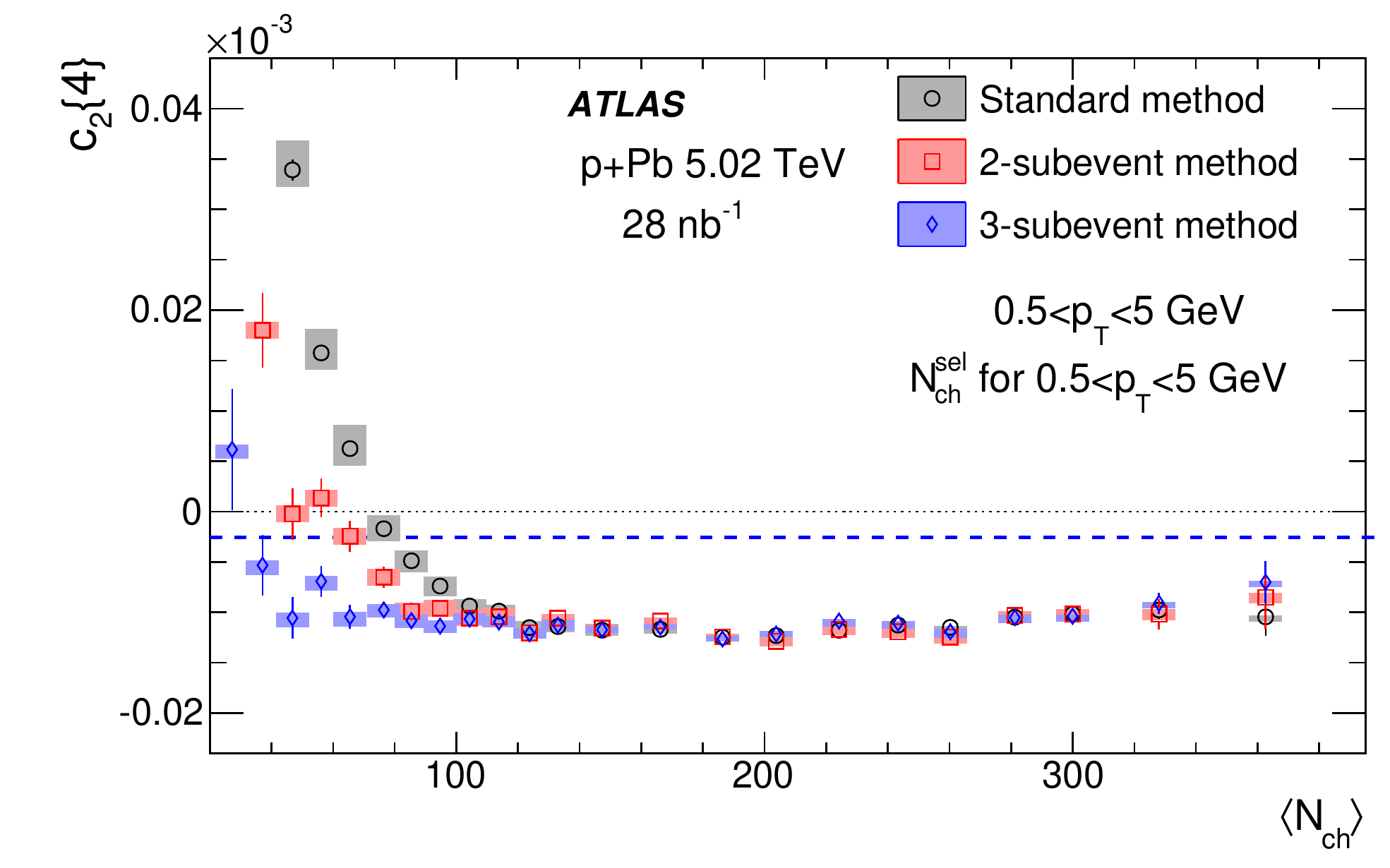}
\end{center}\vspace*{-0.6cm}
\caption{\label{fig:6} The $c_2\{4\}$ values calculated for charged particles with $0.3<\pT<3$~GeV (left panel) and $0.5<\pT<5$~GeV (right panel) compared for the three cumulant methods from the 5.02~TeV $\pPb$ data. The event averaging is performed for $\nchb$ calculated for the same $\pT$ range, which is then mapped to $\nch$, the average number of charged particles with $\pT>0.4$~GeV. The dashed line indicates the $c_2\{4\}$ value corresponding to a 4\% $v_2$ signal. The error bars and shaded boxes represent the statistical and systematic uncertainties, respectively.}
\end{figure}

Figures~\ref{fig:4}--\ref{fig:6} show direct comparisons of the results for the standard, two-subevent, and three-subevent methods for $\pp$ collisions at $\sqrt{s}=13$~TeV, $\pp$ at $\sqrt{s}=5.02$~TeV and $\pPb$ collisions at $\sqrtsnn=5.02$~TeV, respectively. The results from 5.02~TeV $\pp$ collisions are qualitatively similar to those from the 13~TeV $\pp$ collisions, i.e.\ the $c_{2}\{4\}$ values are smallest for the three-subevent method and largest for the standard method. The same hierarchy between the three methods is also observed in $\pPb$ collisions, but only for the $\nch<100$ region, suggesting that non-flow effects in $\pPb$ collisions are much smaller than those in $\pp$ collisions at comparable $\nch$. In $\pPb$ collisions, all three methods give consistent results for $\nch>100$. Furthermore, the three-subevent method gives negative $c_2\{4\}$ values in most of the measured $\nch$ range. 

\begin{figure}[!h]
\begin{center}
\includegraphics[width=0.9\linewidth]{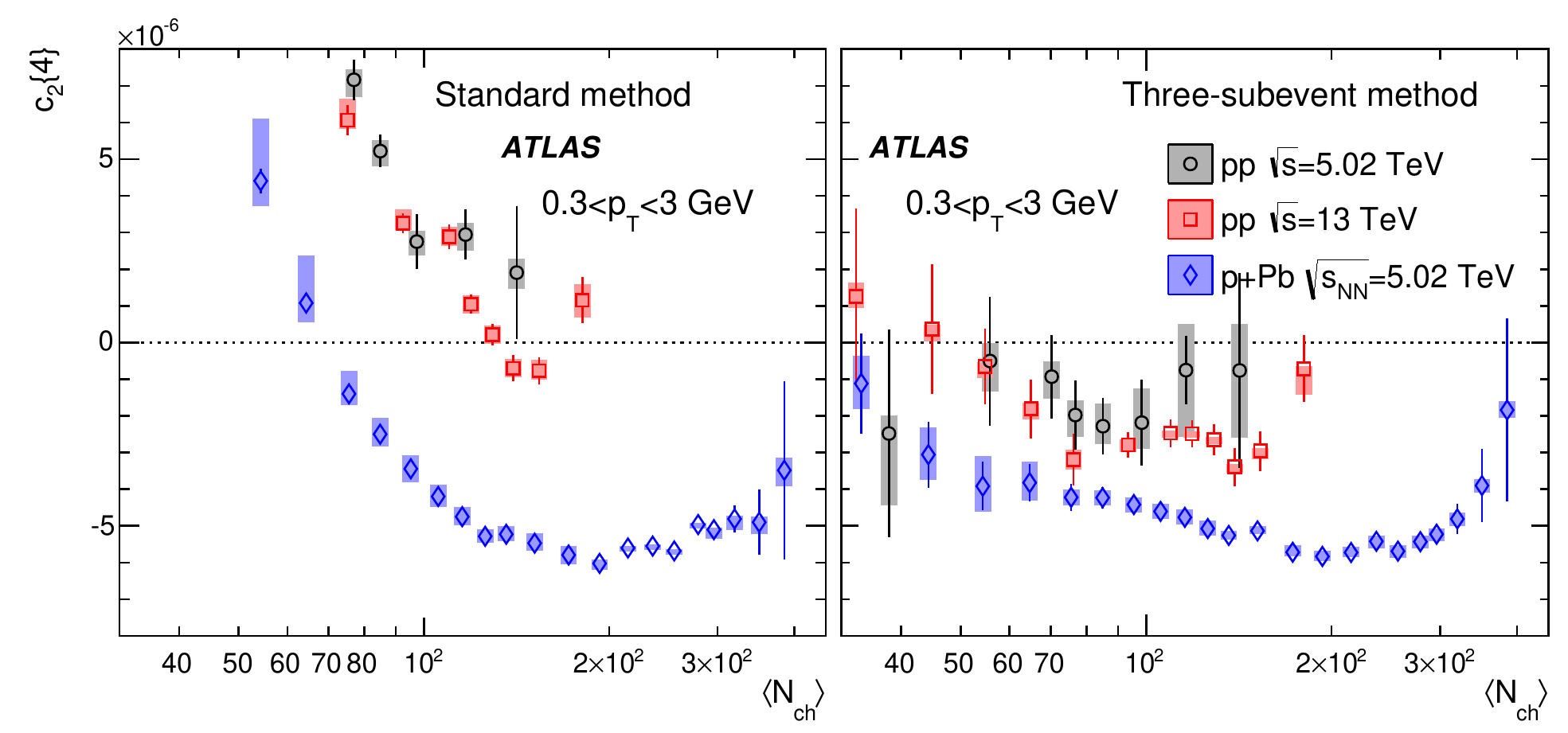}
\end{center}
\caption{\label{fig:7}  The $c_2\{4\}$ values calculated for charged particles with $0.3<\pT<3$~GeV using the standard cumulants (left panel) and the three-subevent method (right panel) compared between 5.02~TeV $\pp$, 13~TeV $\pp$ and 5.02~TeV $\pPb$. The event averaging is performed for $\nchb$ calculated for the same $\pT$ range, which is then mapped to $\nch$, the average number of charged particles with $\pT>0.4$~GeV. The error bars and shaded boxes represent the statistical and systematic uncertainties, respectively.}
\end{figure}

\begin{figure}[!h]
\begin{center}
\includegraphics[width=0.9\linewidth]{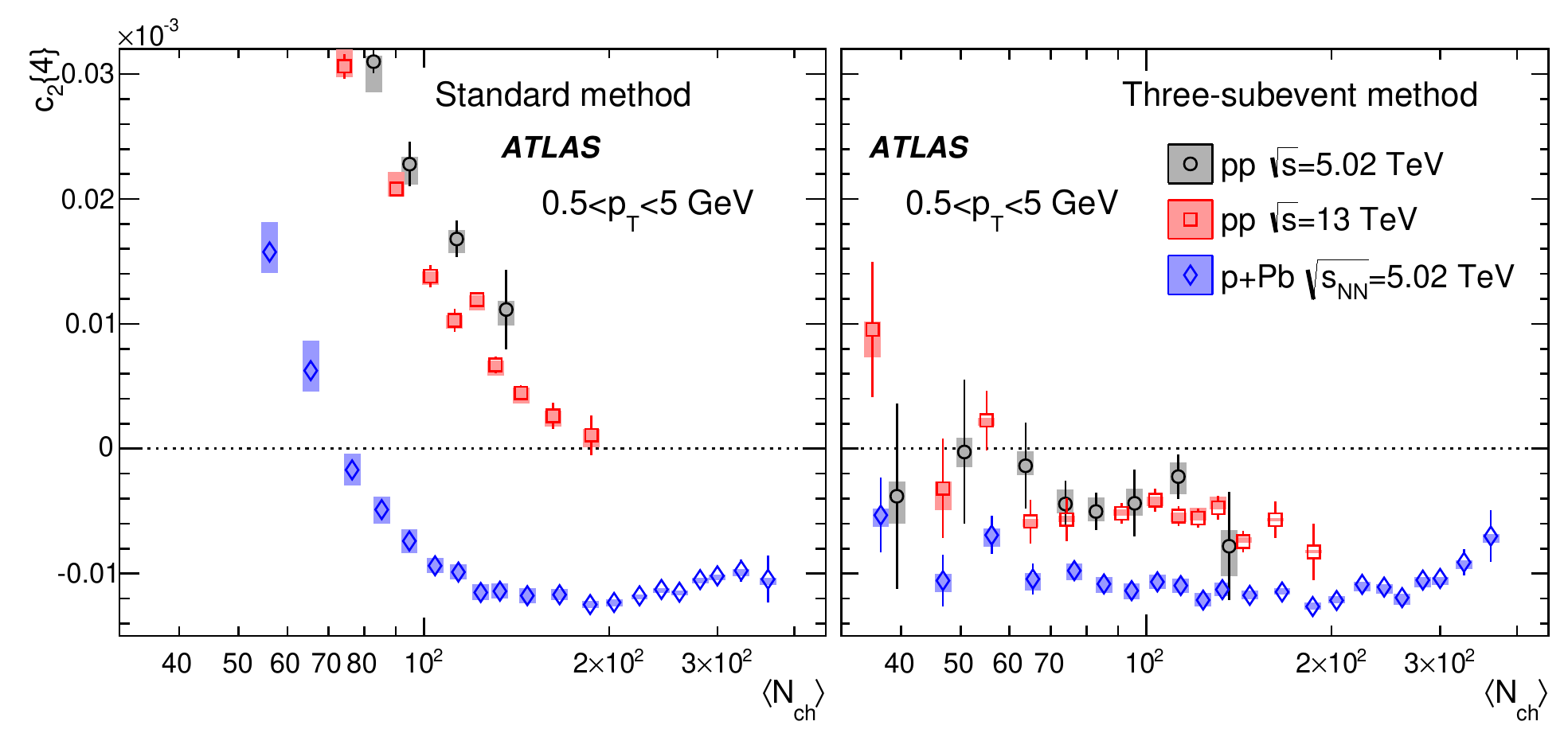}
\end{center}
\caption{\label{fig:8}  The $c_2\{4\}$ values calculated for charged particles with $0.5<\pT<5$~GeV using the standard cumulants (left panel) and the three-subevent method (right panel) compared between 5.02~TeV $\pp$, 13~TeV $\pp$ and 5.02~TeV $\pPb$. The event averaging is performed for $\nchb$ calculated for the same $\pT$ range, which is then mapped to $\nch$, the average number of charged particles with $\pT>0.4$~GeV. The error bars and shaded boxes represent the statistical and systematic uncertainties, respectively.}
\end{figure}

The comparison of the $c_2\{4\}$ values between the three datasets, for the standard and the three-subevent methods, is shown in Figures~\ref{fig:7} and \ref{fig:8}. The large positive $c_2\{4\}$ values observed in the small $\nch$ region in the standard method are likely due to non-flow correlations, since this trend is absent when using the three-subevent cumulant method. In $\pPb$ collisions, the absolute value of $c_2\{4\}$ seems to become smaller for $\nch>200$.


\begin{figure}[!h]
\begin{center}
\includegraphics[width=0.9\linewidth]{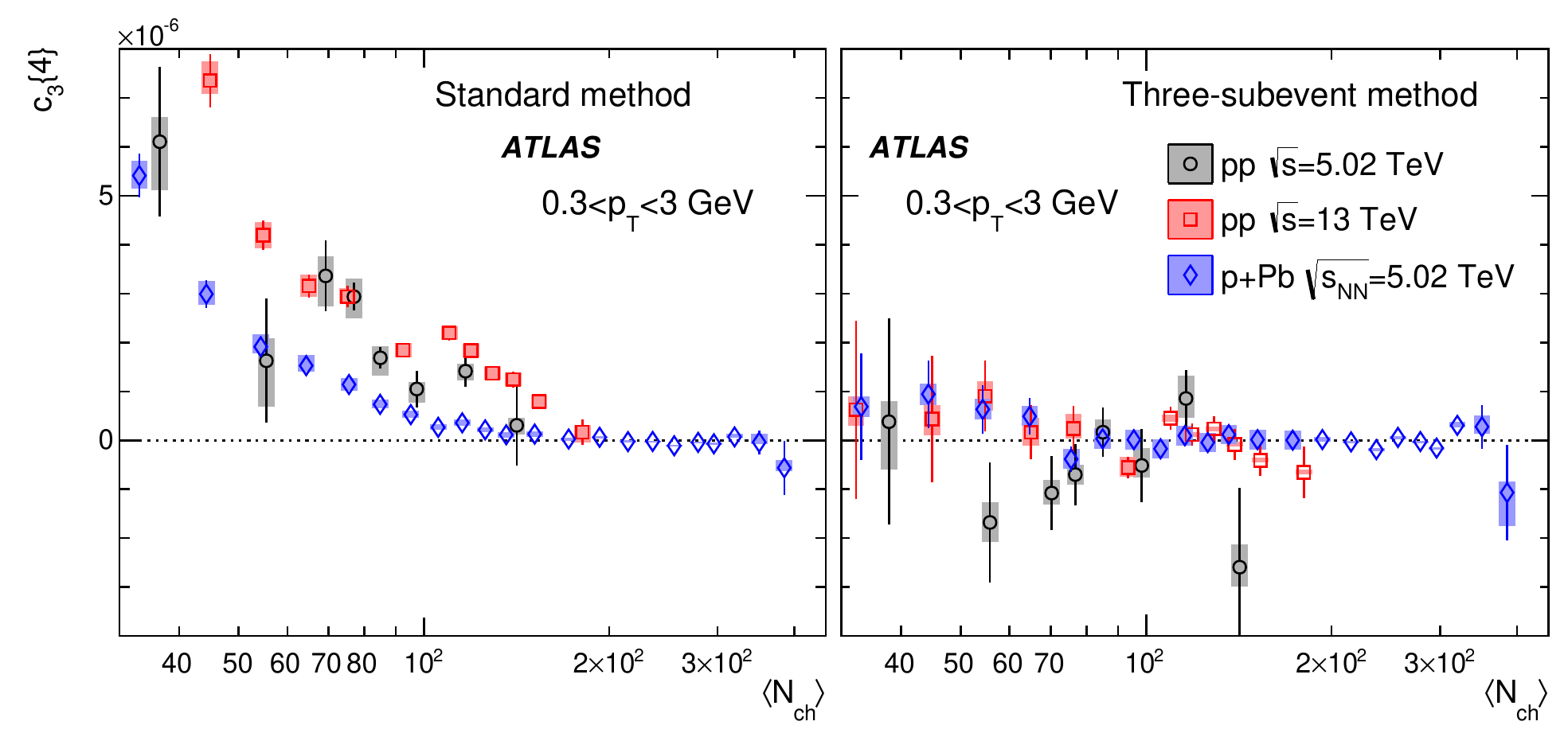}
\end{center}
\caption{\label{fig:9} The $c_3\{4\}$ values calculated for charged particles with $0.3<\pT<3$~GeV using the standard cumulants (left panel) and the three-subevent method (right panel) compared between 5.02~TeV $\pp$, 13~TeV $\pp$ and 5.02~TeV $\pPb$. The event averaging is performed for $\nchb$ calculated for the same $\pT$ range, which is then mapped to $\nch$, the average number of charged particles with $\pT>0.4$~GeV. The error bars and shaded boxes represent the statistical and systematic uncertainties, respectively.}
\end{figure}

\begin{figure}[!h]
\begin{center}
\includegraphics[width=0.9\linewidth]{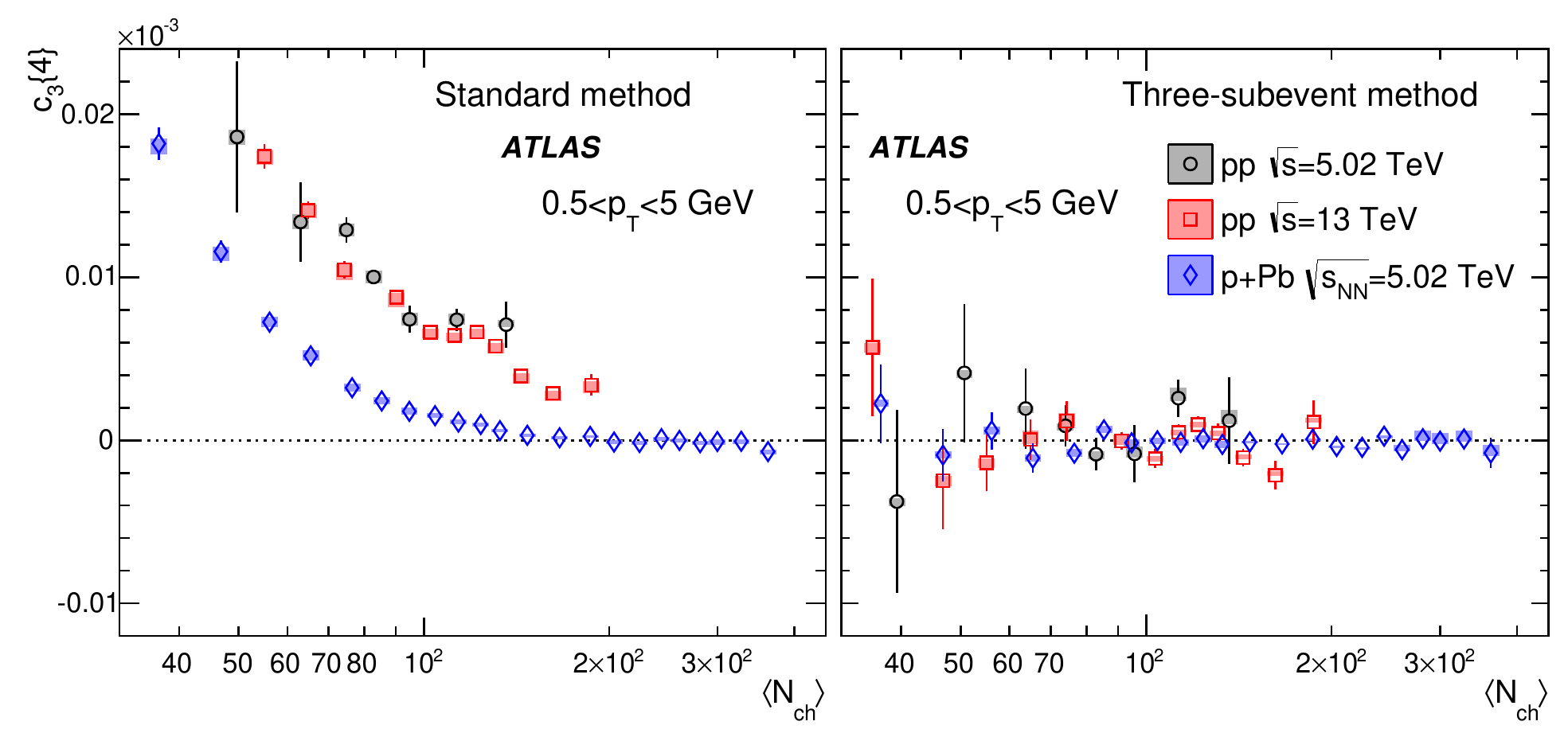}
\end{center}
\caption{\label{fig:10}  The $c_3\{4\}$ values calculated for charged particles with $0.5<\pT<5$~GeV using the standard cumulants (left panel) and the three-subevent method (right panel) compared between 5.02~TeV $\pp$, 13~TeV $\pp$ and 5.02~TeV $\pPb$. The event averaging is performed for $\nchb$ calculated for the same $\pT$ range, which is then mapped to $\nch$, the average number of charged particles with $\pT>0.4$~GeV. The error bars and shaded boxes represent the statistical and systematic uncertainties, respectively.}
\end{figure}

The same analysis is performed for the third-order harmonics. Figures~\ref{fig:9} and \ref{fig:10} compare the $c_3\{4\}$ values between the three datasets for the standard cumulant method and the three-subevent method. The $c_3\{4\}$ values from the three-subevent method are close to zero in all three systems. For the standard method, the positive $c_3\{4\}$ values in the small $\nch$ region indicate the influence of non-flow correlations, but the influence is not as strong as that for $c_2\{4\}$.

\begin{figure}[h!]
\begin{center}
\includegraphics[width=0.5\linewidth]{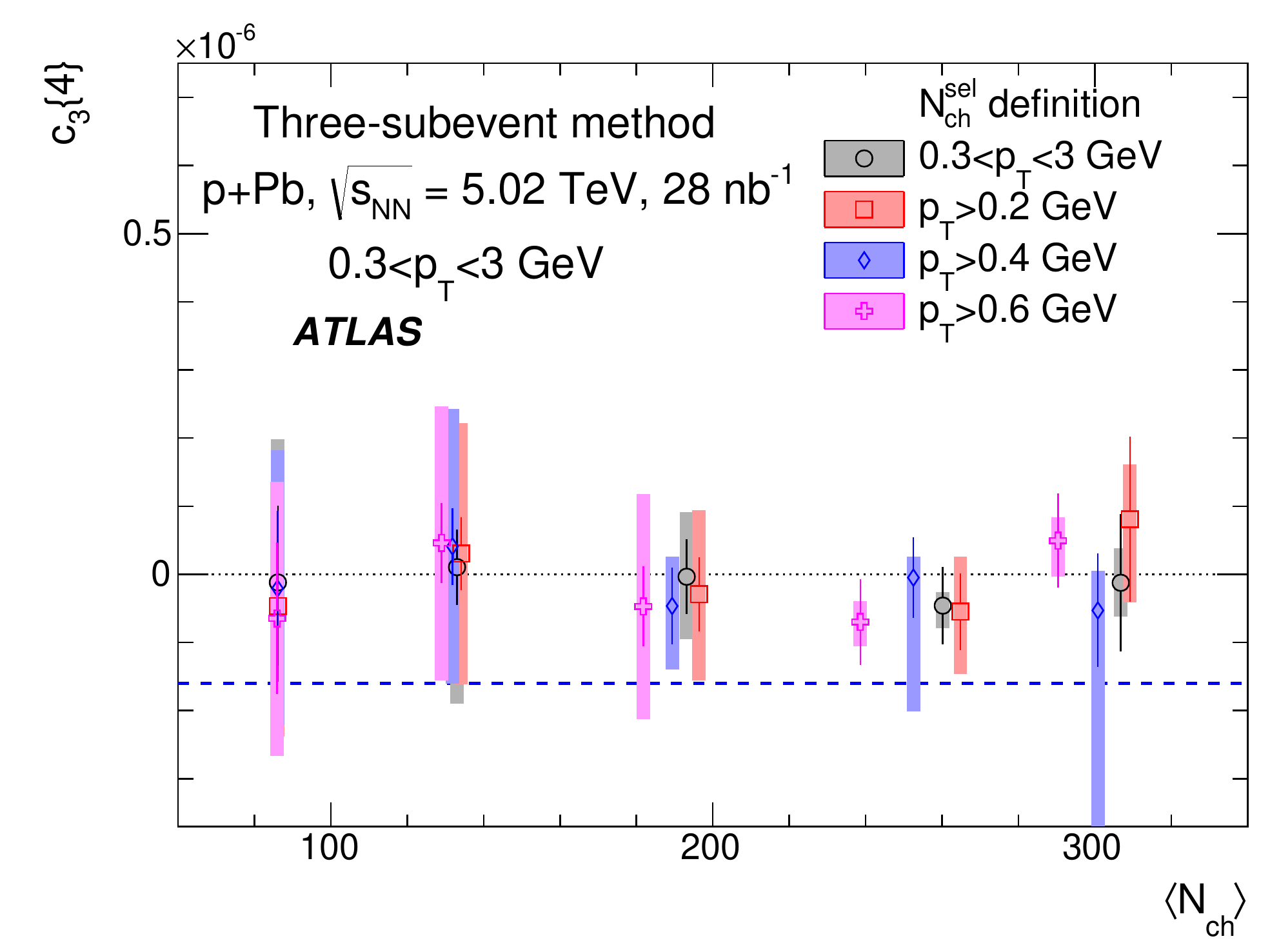}\includegraphics[width=0.5\linewidth]{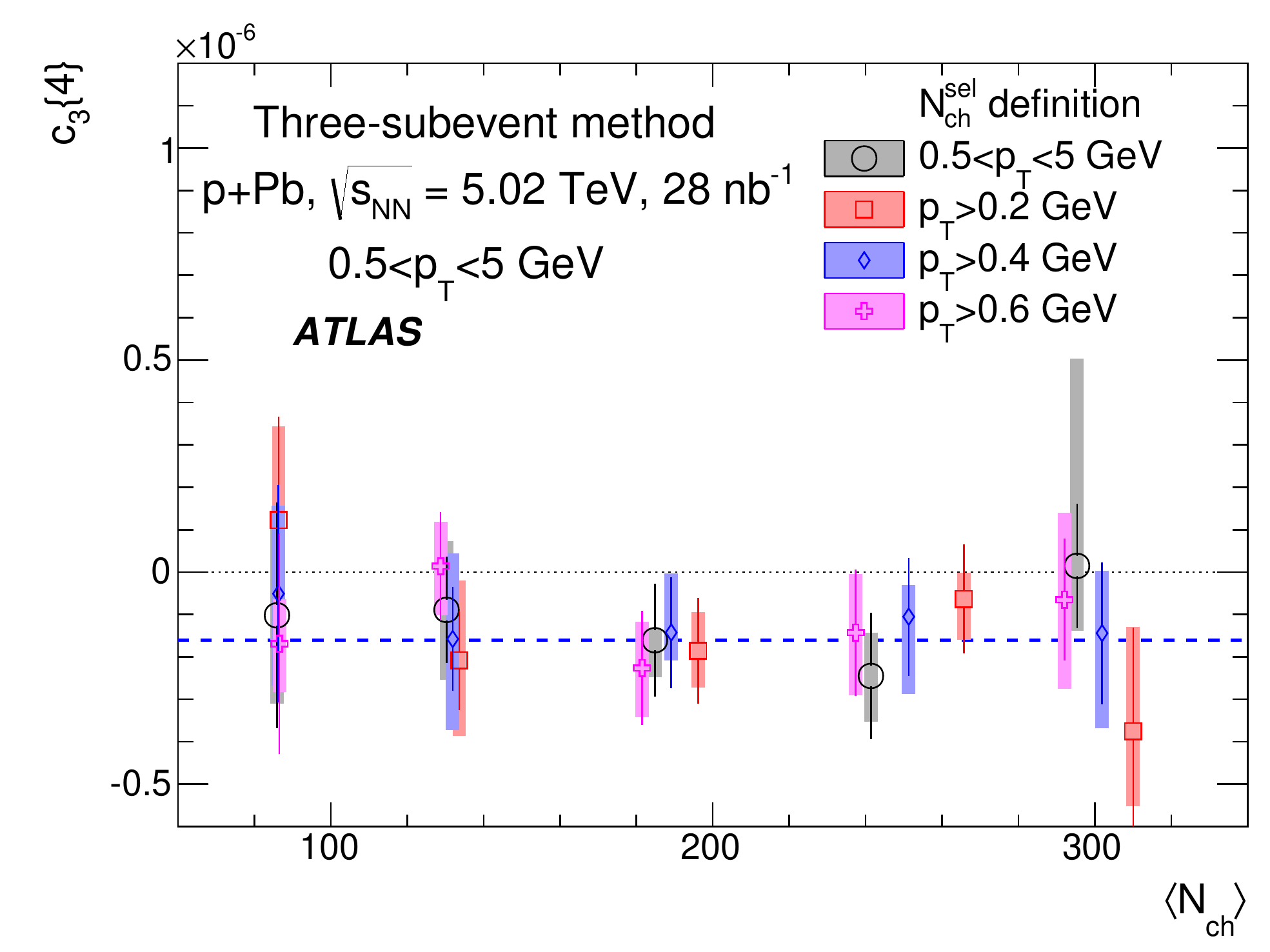}
\end{center}
\caption{\label{fig:10b} The $c_3\{4\}$ values calculated for charged particles with $0.3<\pT<3$~GeV (left panel) or $0.5<\pT<5$~GeV (right panel) with the three-subevent cumulant method for the $\pPb$ data. The event averaging is performed for $\nchb$ calculated for various $\pT$ selections as indicated in the figure, which is then mapped to $\nch$, the average number of charged particles with $\pT>0.4$~GeV. The dashed line indicates the $c_3\{4\}$ value corresponding to a 2\% $v_3$ signal. The error bars and shaded boxes represent the statistical and systematic uncertainties, respectively.}
\end{figure}

Figure~\ref{fig:10b} shows the $c_3\{4\}$ values from $\pPb$ collisions in the two $\pT$ ranges, obtained with the three-subevent method; they are zoomed-in version of the $\pPb$ data shown in Figures~\ref{fig:8} and \ref{fig:9}. Within their large statistical and systematic uncertainties, the values of $c_3\{4\}$ are systematically below zero, especially for $0.5<\pT<5$~GeV, where the $c_3\{4\}$ values are comparable to $-0.16\times10^{-6}$, corresponding to a $v_3$ value of 2\% as indicated in the figure. The negative $c_3\{4\}$ values from the three-subevent method support the existance of long-range multi-particle triangular flow in $\pPb$ collisions. 

\subsection{Three-subevent flow harmonic $v_2\{4\}$}
\begin{figure}[!h]
\begin{center}
\includegraphics[width=1\linewidth]{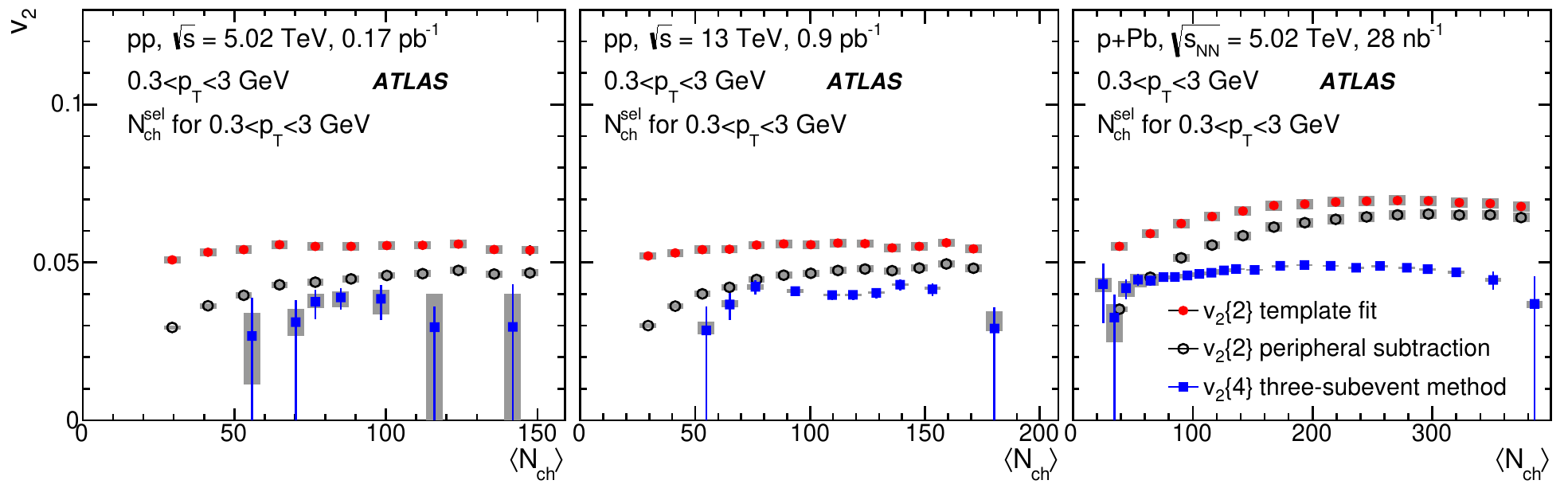}
\end{center}
\caption{\label{fig:11} The $v_2\{4\}$ values calculated for charged particles with $0.3<\pT<3$~GeV using the three-subevent method in 5.02~TeV $\pp$ (left panel), 13~TeV $\pp$ (middle panel) and 5.02~TeV $\pPb$ collisions (right panel). They are compared to $v_2$ obtained from the 2PC analyses~\cite{Aad:2014lta,Aaboud:2016yar} where the non-flow effects are removed by a template fit procedure (solid circles) or with a fit after subtraction with a ZYAM assumption (peripheral subtraction, open circles). The error bars and shaded boxes represent the statistical and systematic uncertainties, respectively.}
\end{figure}

\begin{figure}[!h]
\begin{center}
\includegraphics[width=1\linewidth]{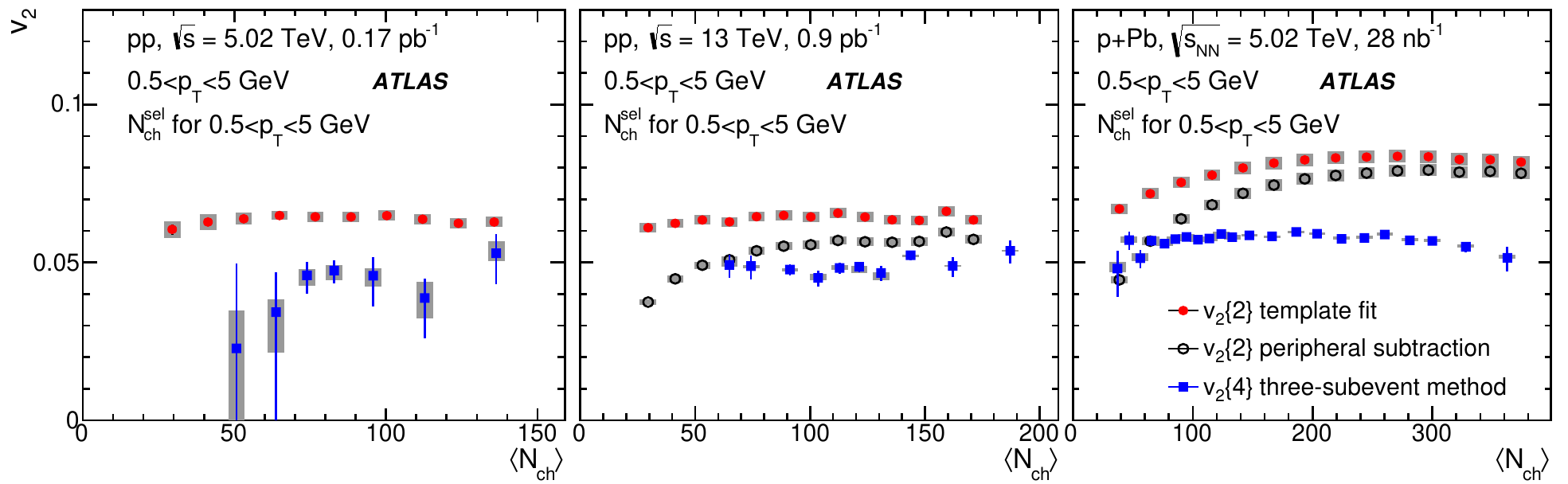}
\end{center}
\caption{\label{fig:12} The $v_2\{4\}$ values calculated for charged particles with $0.5<\pT<5$~GeV using the three-subevent method in 5.02~TeV $\pp$ (left panel), 13~TeV $\pp$ (middle panel) and 5.02~TeV $\pPb$ collisions (right panel). They are compared to $v_2$ obtained from the 2PC analyses~\cite{Aad:2014lta,Aaboud:2016yar} where the non-flow effects are removed by a template fit procedure (solid circles) or with a fit after subtraction with a ZYAM assumption (peripheral subtraction, open circles). The error bars and shaded boxes represent the statistical and systematic uncertainties, respectively.}
\end{figure}

The harmonic flow coefficients $v_2\{4\}$ can be obtained from the measured values of $c_2\{4\}$ according to Eq.~\eqref{eq:6}. Figure~\ref{fig:11} shows the $v_2\{4\}$ values for charged particles with $0.3<\pT<3$~GeV calculated using the three-subevent method in the three datasets. Results for the higher $\pT$ range ($0.5<\pT<5$~GeV) are presented in Figure~\ref{fig:12}. The value of $v_2\{4\}$ is measured down to $\nch\approx50$ in $\pp$ collisions and down to $\nch\approx 20$--40 in $\pPb$ collisions. The $v_2\{4\}$ values are observed to be approximately independent of $\nch$ in the measured range in the three datasets: $50<\nch<150$ for 5.02~TeV $\pp$, $50<\nch<200$ for 13~TeV $\pp$, and $20<\nch<380$ for 5.02~TeV $\pPb$, respectively. Moreover, the $\pPb$ data suggest the value of $v_2\{4\}$ is lower for $\nch>200$, as expected from the similar behavior of $|c_2\{4\}|$ in Figures~\ref{fig:7} and \ref{fig:8} at large $\nch$.

The values of $v_2\{4\}$ presented in Figures~\ref{fig:11} and \ref{fig:12} are also compared to the values of $v_2\{2\}$ obtained from the 2PC measurements~\cite{Aad:2014lta,Aaboud:2016yar} where the non-flow effects are estimated using low-multiplicity events ($\nch<20$) and then subtracted. The subtraction was performed either by a ``template fit'', which includes the pedestal level from the $\nch<20$ events, or by a ``peripheral subtraction'', which sets the pedestal level by a zero-yield at minimum (ZYAM) procedure~\cite{Adare:2008ae}. The ``peripheral subtraction'' explicitly assumes that the most peripheral events do not contain any long-range correlations~\cite{Aaboud:2016yar}, and so $v_2$ is forced to be zero at the corresponding $\nch$ value, which biases $v_2$ to a lower value in other multiplicity ranges.

\begin{figure}[!h]
\begin{center}
\includegraphics[width=0.5\linewidth]{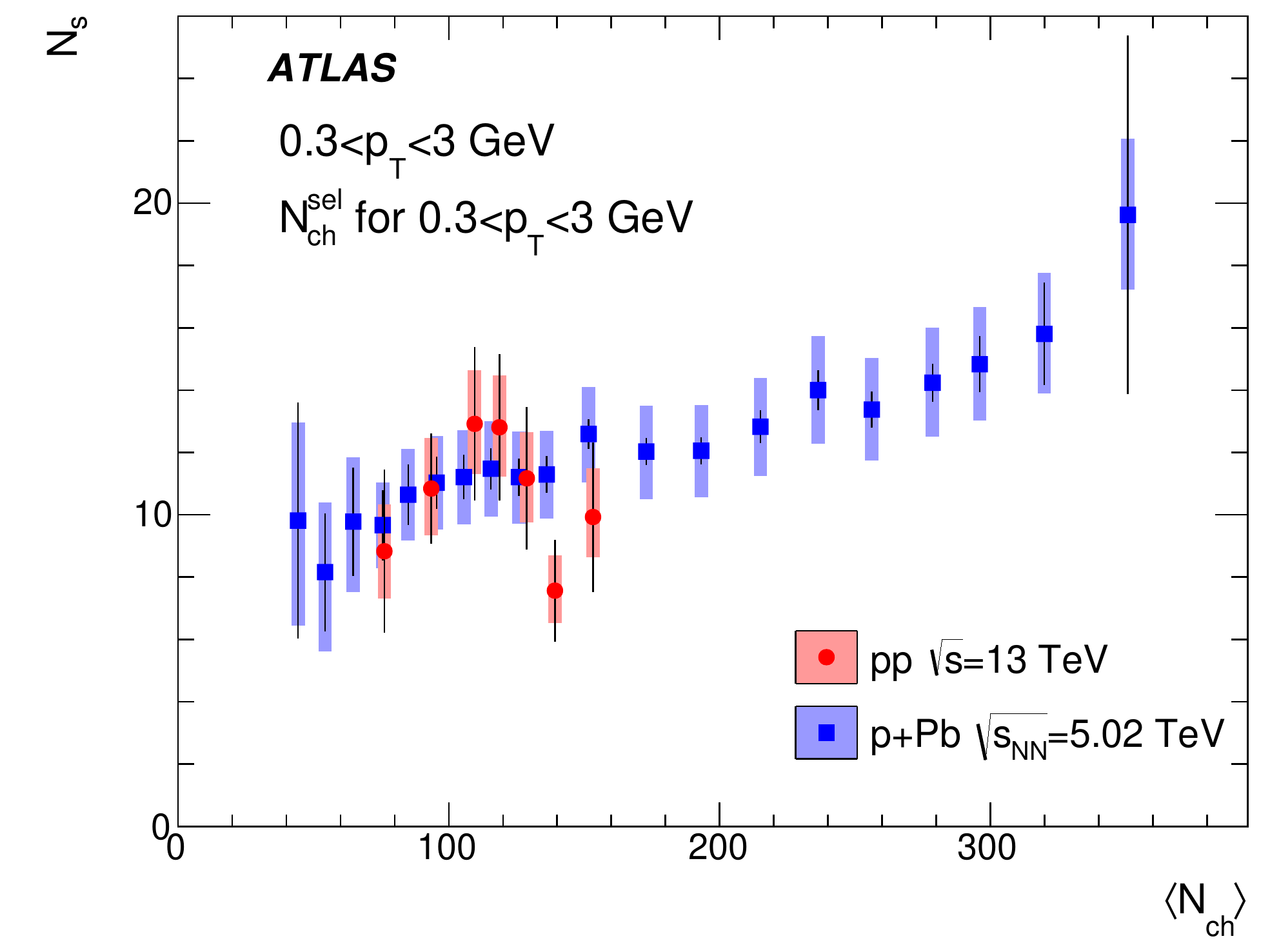}\includegraphics[width=0.5\linewidth]{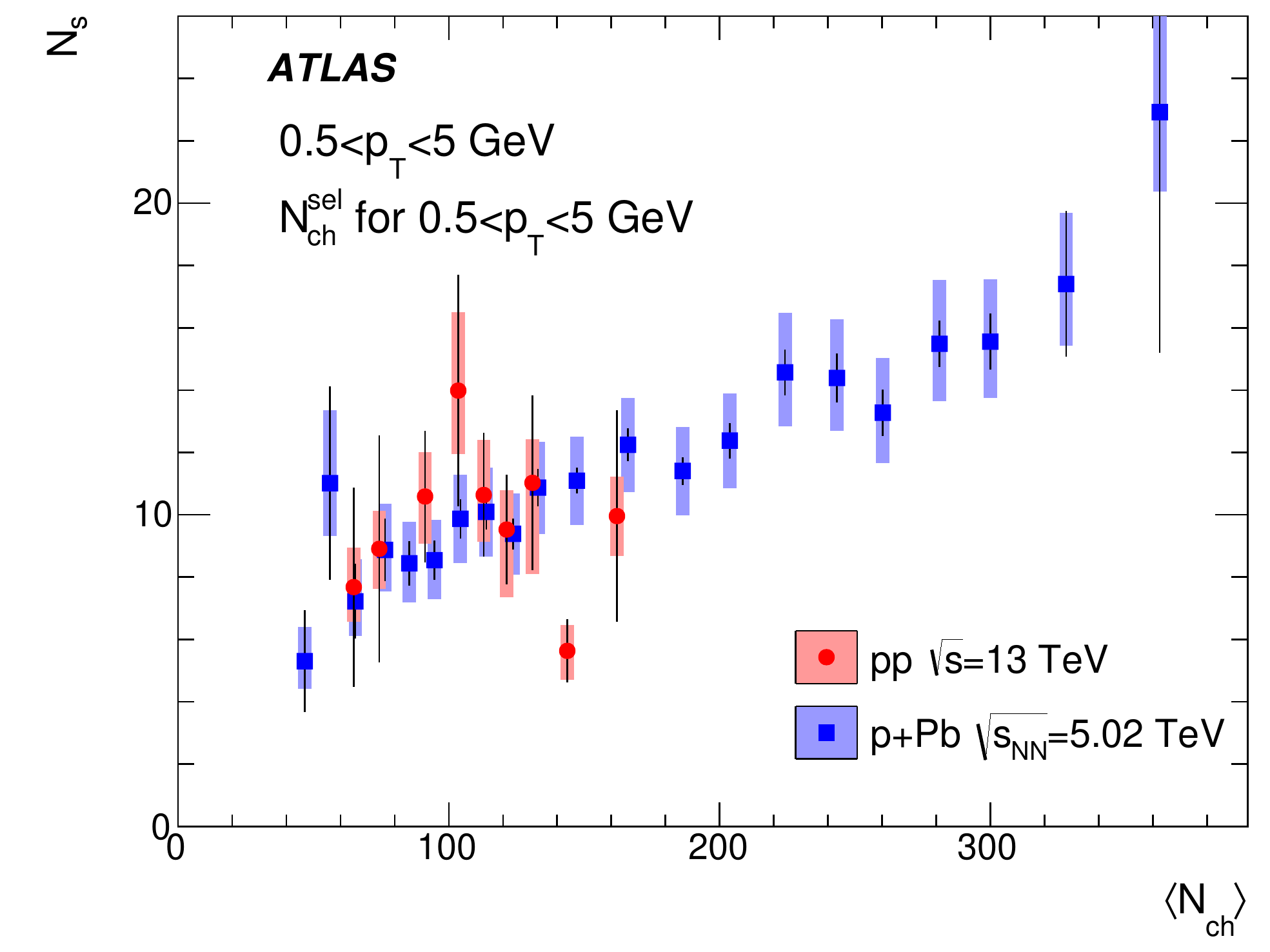}
\end{center}
\caption{\label{fig:13} The number of sources inferred from $v_2\{2\}$ and $v_2\{4\}$ measurements via the model framework in Refs.~\cite{Bzdak:2013rya,Yan:2013laa} and Eq.~\eqref{eq:13} in 13~TeV $\pp$ and 5.02~TeV $\pPb$ collisions, for charged particles with $0.3<\pT<3$~GeV (left panel) and $0.5<\pT<5$~GeV (right panel). The error bars and shaded boxes represent the statistical and systematic uncertainties, respectively.}
\end{figure}

\subsection{Dependence on the number of sources in the initial state}
Figures~\ref{fig:11} and \ref{fig:12} show that the $v_2\{4\}$ values are smaller than the $v_2\{2\}$ values extracted using the template-fit method in both the $\pp$ and $\pPb$ collisions. In various hydrodynamic models for small collision systems~\cite{Bzdak:2013rya,Yan:2013laa}, this difference can be interpreted as the influence of event-by-event flow fluctuations associated with the initial state, which is closely related to the effective number of sources $N_{\mathrm{s}}$ for particle production in the transverse density distribution of the initial state~\cite{Yan:2013laa}:
\begin{eqnarray}
\label{eq:13}
\frac{v_2\{4\}}{v_2\{2\}} = \left[\frac{4}{(3+N_{\mathrm{s}})}\right]^{1/4}\;\;\mbox{or}\;\; N_{\mathrm{s}} = \frac{4v_2\{2\}^4}{v_2\{4\}^4}-3\;.
\end{eqnarray}

Figure~\ref{fig:13} shows the extracted values of $N_{\mathrm{s}}$ as a function of $\nch$ in 13~TeV $\pp$ and 5.02 $\pPb$ collisions, estimated using charged particles with $0.3<\pT<3$~GeV and $0.5<\pT<5$~GeV. It is observed that the $N_{\mathrm{s}}$ value increases with $\nch$ in $\pPb$ collisions, reaching $N_{\mathrm{s}}\sim20$ in the highest multiplicity class, and it is consistent between the two $\pT$ ranges. 

In the model framework in Refs.~\cite{Bzdak:2013rya,Yan:2013laa}, the values of $|c_2\{4\}|$ and $v_2\{4\}$ are expected to decrease for large $N_{\mathrm{s}}$, which is compatible with the presented results. The slight decreases of $|c_2\{4\}|$ shown in Figures~\ref{fig:7} and \ref{fig:8} for $\pPb$ collisions are compatible with the model predictions. The results for 13~TeV $\pp$ collisions cover a limited $\nch$ range compared to $\pPb$, but agree with $\pPb$ collisions in this range.  

\section{Summary}
\label{sec:sum}
Measurements of the four-particle cumulants $c_{n}\{4\}$ and harmonic flow coefficients $v_n\{4\}$ for $n=2$ and 3 are presented using 0.17 pb$^{-1}$ of $\pp$ data at $\sqrt{s}=5.02$~TeV, 0.9 pb$^{-1}$ of $\pp$ data at $\sqrt{s}=13$~TeV and 28 nb$^{-1}$ $\pPb$ of data at $\sqrt{s_{\rm{NN}}}=5.02$~TeV. These measurements were performed with the ATLAS detector at the LHC. The $c_{2}\{4\}$ values are calculated using the standard cumulant method and the recently proposed two-subevent and three-subevent methods. They are all presented as a function of the average number of charged particles with $\pT > 0.4$~GeV, $\nch$. It is found that the $c_{2}\{4\}$ value from the standard method is sensitive to the choice of particles used to form the event classes used for averaging. This suggests that the previous $c_{2}\{4\}$ measurement in $\pp$ collisions~\cite{Khachatryan:2016txc,atlas:4}, based on the standard method, may be dominated by non-flow correlations instead of a long-range collective flow correlation. In general, it is easy to obtain incorrect results from the standard cumulant method, depending on the nature of the non-flow fluctuations associated with the event class chosen for the analysis. 

On the other hand, the sensitivity of $c_{2}\{4\}$ on event class definition is greatly reduced in the two-subevent method and is almost fully removed in the three-subevent method, demonstrating that the three-subevent method is more robust against non-flow effects. Similarly, the values of $c_3\{4\}$ are found to differ in the three datasets using the standard method, but are consistent with each other and much closer to zero using the three-subevent method. This gives confidence that non-flow correlations make a much smaller contribution to the three-subevent results, and that this method is more appropriate for studying long-range collective behaviour than the standard cumulant method.


The three-subevent method provides a measurement of $c_{2}\{4\}$ that is negative in all three datasets over a broad range of $\nch$. The magnitude of $c_{2}\{4\}$ increases with $\pT$ and is nearly independent of $\nch$ but in $\pPb$ collisions the values become smaller at high multiplicities. These results provide direct evidence for the presence of long-range multi-particle azimuthal correlations in broad $\nch$ ranges in $\pp$ and $\pPb$ collisions, and these long-range multi-particle correlations persist even in events with rather low multiplicity of $\nch\sim40$. The $c_{3}\{4\}$ values are consistent with zero in $\pp$ collisions, but are systematically below zero in $\pPb$ collisions, compatible with the presence of significant long-range multi-particle triangular flow in $\pPb$ collisions.

The single-particle harmonic coefficient $v_2\{4\}=(-c_{2}\{4\})^{1/4}$ is calculated and compared with $v_2\{2\}$ obtained previously using the two-particle correlation method, where the non-flow contributions were estimated and subtracted. The magnitude of $v_2\{4\}$ is smaller than that for $v_2\{2\}$, as expected for a long-range final-state hydrodynamic collective effect. The ratio of $v_2\{4\}$ to $v_2\{2\}$ is used, in a model-dependent framework, to infer the number of particle-emitting sources in the initial-state geometric configuration. The number of sources extracted within this framework is found to increase with $\nch$ in $\pPb$ collisions.


The subevent cumulant technique and the new results provide direct evidence that the ridge is indeed a long-range collective phenomenon involving many particles distributed across a broad rapidity interval. The results of $v_2\{4\}$ and its dependence on $\pT$ and $\nch$, largely free from non-flow effects, can be used to understand the space-time dynamics and the properties of the medium created in small collision systems.

\section*{Acknowledgments}

We thank CERN for the very successful operation of the LHC, as well as the
support staff from our institutions without whom ATLAS could not be
operated efficiently.

We acknowledge the support of ANPCyT, Argentina; YerPhI, Armenia; ARC, Australia; BMWFW and FWF, Austria; ANAS, Azerbaijan; SSTC, Belarus; CNPq and FAPESP, Brazil; NSERC, NRC and CFI, Canada; CERN; CONICYT, Chile; CAS, MOST and NSFC, China; COLCIENCIAS, Colombia; MSMT CR, MPO CR and VSC CR, Czech Republic; DNRF and DNSRC, Denmark; IN2P3-CNRS, CEA-DSM/IRFU, France; SRNSF, Georgia; BMBF, HGF, and MPG, Germany; GSRT, Greece; RGC, Hong Kong SAR, China; ISF, I-CORE and Benoziyo Center, Israel; INFN, Italy; MEXT and JSPS, Japan; CNRST, Morocco; NWO, Netherlands; RCN, Norway; MNiSW and NCN, Poland; FCT, Portugal; MNE/IFA, Romania; MES of Russia and NRC KI, Russian Federation; JINR; MESTD, Serbia; MSSR, Slovakia; ARRS and MIZ\v{S}, Slovenia; DST/NRF, South Africa; MINECO, Spain; SRC and Wallenberg Foundation, Sweden; SERI, SNSF and Cantons of Bern and Geneva, Switzerland; MOST, Taiwan; TAEK, Turkey; STFC, United Kingdom; DOE and NSF, United States of America. In addition, individual groups and members have received support from BCKDF, the Canada Council, CANARIE, CRC, Compute Canada, FQRNT, and the Ontario Innovation Trust, Canada; EPLANET, ERC, ERDF, FP7, Horizon 2020 and Marie Sk{\l}odowska-Curie Actions, European Union; Investissements d'Avenir Labex and Idex, ANR, R{\'e}gion Auvergne and Fondation Partager le Savoir, France; DFG and AvH Foundation, Germany; Herakleitos, Thales and Aristeia programmes co-financed by EU-ESF and the Greek NSRF; BSF, GIF and Minerva, Israel; BRF, Norway; CERCA Programme Generalitat de Catalunya, Generalitat Valenciana, Spain; the Royal Society and Leverhulme Trust, United Kingdom.

The crucial computing support from all WLCG partners is acknowledged gratefully, in particular from CERN, the ATLAS Tier-1 facilities at TRIUMF (Canada), NDGF (Denmark, Norway, Sweden), CC-IN2P3 (France), KIT/GridKA (Germany), INFN-CNAF (Italy), NL-T1 (Netherlands), PIC (Spain), ASGC (Taiwan), RAL (UK) and BNL (USA), the Tier-2 facilities worldwide and large non-WLCG resource providers. Major contributors of computing resources are listed in Ref.~\cite{ATL-GEN-PUB-2016-002}.

\bibliographystyle{atlasBibStyleWithTitle}
\bibliography{hion-2017-02}


\clearpage

\newpage 
\begin{flushleft}
{\Large The ATLAS Collaboration}

\bigskip

M.~Aaboud$^\textrm{\scriptsize 137d}$,
G.~Aad$^\textrm{\scriptsize 88}$,
B.~Abbott$^\textrm{\scriptsize 115}$,
O.~Abdinov$^\textrm{\scriptsize 12}$$^{,*}$,
B.~Abeloos$^\textrm{\scriptsize 119}$,
S.H.~Abidi$^\textrm{\scriptsize 161}$,
O.S.~AbouZeid$^\textrm{\scriptsize 139}$,
N.L.~Abraham$^\textrm{\scriptsize 151}$,
H.~Abramowicz$^\textrm{\scriptsize 155}$,
H.~Abreu$^\textrm{\scriptsize 154}$
% ...
\bigskip
\\
% ...
$^{12}$ Institute of Physics, Azerbaijan Academy of Sciences, Baku, Azerbaijan\\
% ...
$^{88}$ CPPM, Aix-Marseille Universit{\'e} and CNRS/IN2P3, Marseille, France\\
% ...
$^{115}$ Homer L. Dodge Department of Physics and Astronomy, University of Oklahoma, Norman OK, United States of America\\
% ...
$^{119}$ LAL, Univ. Paris-Sud, CNRS/IN2P3, Universit{\'e} Paris-Saclay, Orsay, France\\
% ...
$^{137}$ $^{(a)}$ Facult{\'e} des Sciences Ain Chock, R{\'e}seau Universitaire de Physique des Hautes Energies - Universit{\'e} Hassan II, Casablanca; $^{(b)}$ Centre National de l'Energie des Sciences Techniques Nucleaires, Rabat; $^{(c)}$ Facult{\'e} des Sciences Semlalia, Universit{\'e} Cadi Ayyad, LPHEA-Marrakech; $^{(d)}$ Facult{\'e} des Sciences, Universit{\'e} Mohamed Premier and LPTPM, Oujda; $^{(e)}$ Facult{\'e} des sciences, Universit{\'e} Mohammed V, Rabat, Morocco\\
% ...
$^{139}$ Santa Cruz Institute for Particle Physics, University of California Santa Cruz, Santa Cruz CA, United States of America\\
% ...
$^{151}$ Department of Physics and Astronomy, University of Sussex, Brighton, United Kingdom\\
% ...
$^{154}$ Department of Physics, Technion: Israel Institute of Technology, Haifa, Israel\\
$^{155}$ Raymond and Beverly Sackler School of Physics and Astronomy, Tel Aviv University, Tel Aviv, Israel\\
% ...
$^{161}$ Department of Physics, University of Toronto, Toronto ON, Canada\\
% ...
$^{a}$ Also at Department of Physics, King's College London, London, United Kingdom\\
$^{b}$ Also at Institute of Physics, Azerbaijan Academy of Sciences, Baku, Azerbaijan\\
$^{c}$ Also at Novosibirsk State University, Novosibirsk, Russia\\
$^{d}$ Also at TRIUMF, Vancouver BC, Canada\\
$^{e}$ Also at Department of Physics {\&} Astronomy, University of Louisville, Louisville, KY, United States of America\\
% ...
$^{*}$ Deceased
\end{flushleft}

 
\end{document}